\documentclass[11pt]{article}

\usepackage{cite}
 \usepackage{subfigure}
\usepackage{multirow}
\usepackage{helvet}
\usepackage{amsmath}
\usepackage{amssymb}
\usepackage{setspace}
\usepackage{graphicx}
\usepackage{empheq}
\usepackage{soul}
\usepackage{tabularx}
\usepackage{array}
\usepackage{float}
\usepackage{braket}
\usepackage[utf8]{inputenc}
\usepackage{url}
\usepackage{hyperref}
\usepackage{slashed}
\usepackage[font=footnotesize,labelfont=bf]{caption}
\usepackage{color}

\def\be{\begin{equation}}
\def\ee{\end{equation}}

\usepackage[a4paper,top=3cm,bottom=3cm,left=2.5cm,right=2.5cm,bindingoffset=0mm]{geometry}

\begin{document}

\begin{center}

{\Large \bf A Reassessment of the Role of High $x$ Data\\\vspace{0.3cm} on the MSHT Global PDF Fit}

\vspace*{1cm}
L. A. Harland-Lang$^{a}$, T. Cridge$^b$, M. Reader$^a$
and R.S. Thorne$^a$\\                                               
\vspace*{0.5cm}                                                    
$^a$ Department of Physics and Astronomy, University College London, London, WC1E 6BT, UK \\   
$^b$ Elementary Particle Physics, University of Antwerp, Groenenborgerlaan 171, 2020 Antwerp, Belgium 

\begin{abstract}
\noindent We present updates within the MSHT global PDF fit that focus on the high $x$ region, and on improving our understanding of the interplay of various theoretical contributions and experimental constraints here. We revisit the question of target mass and higher twist corrections, considering their impact for the first time at approximate N$^3$LO order in a global PDF analysis. Their inclusion is found to be moderate but not negligible on both the PDFs and preferred value of the strong coupling. Increased  stability in these at aN$^3$LO is observed in comparison to lower orders. We also study the impact of an updated treatment of various fixed--target DIS data, the inclusion of Seaquest fixed--target Drell Yan data, and new ZEUS data that extends coverage into the high $x$ region. The Seaquest data have the largest effect of these, in particular on the light quark separation at high $x$, while the impact of the other updates is rather mild.

\end{abstract}

\end{center}
 
\begin{spacing}{1.2}
\clearpage
\tableofcontents
\clearpage
\end{spacing}

\section{Introduction}

The parton distribution functions (PDFs) of the proton are an essential component of the LHC precision physics program. Various groups work on extracting PDFs as accurately and precisely as possible and release corresponding PDF sets~\cite{Bailey:2020ooq,NNPDF:2021njg,Hou:2019efy,ATLAS:2021vod,Alekhin:2017kpj}.  These analyses utilize an increasingly wide selection of data, including Deep Inelastic Scattering (DIS) data from HERA and fixed target experiments and hadron collider data from the Tevatron and LHC, and compare to predictions made using state--of--the--art theoretical calculations. Next--to--next--to leading order (NNLO) in the QCD perturbative expansion is now the default presentation, but both MSHT  
\cite{McGowan:2022nag} and NNPDF \cite{NNPDF:2024nan} have now provided PDF sets obtained with approximate N$^3$LO (aN$^3$LO) QCD calculations and  QED corrections \cite{Cridge:2023ryv,Barontini:2024dyb}. In both cases theoretical uncertainties associated with the perturbative expansion are also estimated. 

The precision of the constraint from the experimental data and the perturbative calculation is such that it is now necessary to be extremely careful about any remaining sources of uncertainty in the PDF fits. For example, we have considered any limitations in precision and accuracy associated with our fixed parameterisation in 
\cite{Harland-Lang:2024kvt}. However, one of the main regions where the PDF uncertainties are larger, and indeed the confidence with which we can ascertain these uncertainties is more open to question, relates the to quark and antiquark distributions at high $x$ values. These are particularly important, since it is the region of high $x$ that corresponds to high partonic centre of mass energy and hence high particle mass that is so vital for searches for BSM physics at the LHC. However, constraints at very high $x$ come from either very high energy final states at hadron colliders, where by definition the statistics are very low and issues relating to contamination from BSM physics are more relevant (see~\cite{Hammou:2023heg,Shen:2024sci} for recent studies), or from 
deep inelastic scattering (DIS) and fixed target Drell Yan (DY) data. At the HERA collider  high $x$ corrections correspond kinematically to high momentum transfer $Q^2$, and again to low statistics. At fixed target DIS experiments larger amounts of data 
can be obtained, and hence, in principle, higher precision is achievable. However, the region of high $x$ and relatively low $Q^2$ automatically corresponds both to large values of $\alpha_S(Q^2)$, accompanied by 
large powers of $\ln(1-x)$ in the perturbative expansion of DIS cross sections, and to low invariant mass $W^2$. The latter is problematic 
due to an enhancement of nonperturbative corrections to the leading order factorization theorem, in particular target mass corrections (TMCs) and higher twist (HT) corrections, both of which are known to grow quickly for falling $W^2$. 

Usually, the MSHT approach to PDF extraction makes a relatively high data cut on $W^2$, to go along with the standard cut on low $Q^2$, in order to circumvent this problem. In \cite{Martin:2003sk} suitable choices for these cuts were determined to be $W^2 >15$ GeV$^2$ and $Q^2>2$ GeV$^2$ for $F_2(x,Q^2)$, and in \cite{Martin:2009iq} $W^2>25$ GeV$^2$ for $F_3(x,Q^2)$ was taken since renormalon calculations imply higher twists are larger for this quantity \cite{Dokshitzer:1995qm,Dasgupta:1996hh}. An investigation of the potential size of higher twist corrections was made in \cite{Martin:2003sk} and in \cite{Thorne:2014toa}, but with target mass effects implicitly included within the phenomenologically determined higher twist correction. However, the precision expected of the PDFs is now somewhat higher than was required in these earlier studies, as is the precision with which the perturbative calculations are made, given MSHT fits at a${\rm N}^3$LO are now  available~\cite{McGowan:2022nag,Cridge:2023ryv,Cridge:2023ozx}.
Moreover, the data sensitive to the high-$x$ region are now more extensive, and more data tensions are apparent. 

Given this, it is appropriate to update details of the MSHT procedure relating to the quark and antiquark determination at high $x$, and to improve the understanding of the interplay of various theoretical and experimental contributions in this region. This is the aim of this paper. We will in particular   reassess the impact of  TMCs and HTs on the MSHT fit, extending previous analyses and for the first time considering the impact of these corrections in a global PDF fit at up to a${\rm N}^3$LO. In contrast to earlier MSHT studies, we will include an explicit calculation of the TMCs, rather than simply parameterise these with the complete higher twist contribution. We will examine the impact of these corrections with both the baseline and a lower $W^2$ cut; in the baseline case we will find that it is difficult to disentangle agnostic fits to HT corrections from missing higher order QCD effects. In terms of the impact on phenomenology we find an encouraging degree of stability in the resulting PDFs with respect to the baseline fit. The impact of HT corrections (when freely fit) is in addition found to be somewhat lower at aN$^3$LO in comparison to the NNLO case, corresponding to an encouraging increase in stability as the perturbative order is increased.  We will also investigate the impact on the preferred value of the strong coupling, finding that these corrections generally lead to a lowering in the central value, but one that remains within the overall uncertainty as evaluated using the MSHT dynamic tolerance procedure.

We note that all of these 
considerations will be particularly important for the fits to structure function data at the EIC. These data will provide much higher experimental precision at high $x$ and relatively high $Q^2$ compared to existing fixed--target data, and with appropriate cuts will lead to improvements in both high-$x$ PDF determination \cite{AbdulKhalek:2021gbh,AbdulKhalek:2022hcn,Armesto:2023hnw} and the determination of $\alpha_S(M_z^2)$ \cite{Cerci:2023uhu,LHLupcoming}. Data will extend into the region which is sensitive to higher twist and target mass corrections, along with higher perturbative orders, and estimating the true uncertainties on PDFs and $\alpha_S(M_Z^2)$ as well as determining the optimum choice of data cuts and/or theoretical corrections to maximize precision and minimize uncertainty will be vital.  

In addition to the above studies, we also present an update of the fixed target data which are used within the MSHT PDF determination, replacing, where appropriate, the data averaged over different energy runs to obtain structure functions with those
at each different energy expressed in terms of reduced cross sections. In some cases this allows us to take into account the correlations between uncertainties in a more complete fashion. Further to this, we investigate the impact of two new datasets with particular sensitivity to the high $x$ region, namely Seaquest data on fixed--target Drell Yan production, and a more recent analysis of ZEUS data on inclusive DIS that extends to the $x\to 1$ region and applies a finer binning. The Seaquest data are found to provide important constraints on the $\overline{d}/\overline{u}$ ratio (or equivalently difference) at high $x$, although  the tension with the Nusea data is also evident. The impact of the ZEUS data, which we assess via reweighting, is found to be rather mild, in particular when the lower $x$ data are removed to avoid double counting with the existing HERA combination data in the fit.

The article is structured as follows. In Section~\ref{sec:FTupdates}
we present the update of the fixed target data which are used within the MSHT PDF determination. In Section~\ref{sec:ZEUS}, we  examine recent high $x$ DIS data from ZEUS, which has the potential to provide additional constraints on the PDFs in this region. In Section~\ref{sec:SQ} we will examine the impact of  recent Seaquest Drell-Yan asymmetry data, which provide a direct constraint on the difference between the $\bar d$ and $\bar u$ distributions, and which is particularly important at aN$^3$LO. In Section~\ref{sec:TMCS} we will then investigate the inclusion of target mass corrections, phenomenological higher twist terms and the dependence on the cuts used for DIS data. In Section~\ref{sec:fitqual} we focus on the impact on the fit quality and PDFs, while in Section~\ref{sec:alphas} we will investigate the impact that various alternative approaches to our usual procedure have on the determination of the strong coupling constant $\alpha_S(M_Z^2)$, and hence obtain an indication of the theoretical/model uncertainty on this due to uncertainty in the manner of dealing with very high $x$ data.     

\section{Fixed Target Data: Updates}\label{sec:FTupdates}

\begin{table}
\begin{center}
  \scriptsize
  \centering
   \renewcommand{\arraystretch}{1.4}
\begin{tabular}{Xrcccc}\hline 
&Old FT (aN${}^3$LO)  &New FT (aN${}^3$LO) &Old FT (NNLO) &New FT (NNLO)
\\ \hline
BCDMS $p$& 183.4 (1.13)& 360.5 (1.10) &179.1 (1.10) & 358.6 (1.09)\\
BCDMS $d$& 149.0 (0.99)& 251.6 (1.02) &148.8 (0.99) &254.8 (1.04)\\
NMC $p$& 120.5 (0.98)& 383.9 (1.57) &124.2 (1.01) &372.4 (1.53)\\
NMC $d$& 101.1 (0.82)& 326.3 (1.34) &112.6 (0.92) &321.8 (1.32)\\
E665 $p$& 68.1 (1.29)& 75.0 (1.41) & 65.4 (1.23) &70.2 (1.32)\\
E665 $d$& 65.4 (1.23)& 72.0 (1.36) & 60.2 (1.14) &63.7 (1.20)\\
NuTeV $F_2$ & 33.2 (0.63)& 35.6 (0.67) &37.9 (0.71) &37.1 (0.70)\\
NuTeV $F_3$ & 28.0 (0.83)&33.3 (0.79)  &33.4 (0.80) &31.9 (0.76)\\
NMC $n/p$&134.1 (0.91)& 144.4 (0.98) & 135.1 (0.91) &139.0 (0.94)\\
{\bf Fixed Target}&{\bf 1424.8 (0.96)} &{\bf 2201.1 (1.11)} &{\bf 1456.4 (0.98)}&{\bf 2204.5 (1.11)}
\\ \hline
{\bf HERA}&{\bf 1593.3 (1.26)}& {\bf 1626.6 (1.29)}&{\bf 1606.7 (1.27)}&{\bf 1622.4 (1.28)} \\
{\bf Hadron Collider}&{\bf 2391.9 (1.34)}&{\bf 2387.1 (1.33)} &{\bf 2494.8 (1.39)}&{\bf 2510.8 (1.40)}
\\ \hline \hline 
{\bf Global }  &{\bf 5410.0 (1.19)}&{\bf 6214.7 (1.23)} &{\bf 5557.8 (1.22)}&{\bf 6337.7 (1.26)} \\
\hline
\end{tabular}
\end{center}
\caption{\sf $\chi^2$ values for MSHT fits, with the old and new treatment of the fixed target (FT) datasets. The absolute value is given, along with the $\chi^2$  per point in brackets, for the individual fixed target datasets that have been updated, as well as results for the global dataset, and subsets of it.  The old (new) treatment corresponds to 1480 (1983) fixed target data points.}
\label{tab:chi2_FTcomp}
\end{table}

As discussed above, we now update a range of the Fixed Target datasets entering the MSHT20 fit to include a more precise account of the correlated systematic uncertainties and separation of the datasets into different beam energies. In summary, the updated datasets are the BCDMS~\cite{BCDMS:1989qop,BCDMS:1989ggw}, NMC~\cite{NewMuon:1996fwh,NewMuon:1996uwk} and E665~\cite{E665:1996mob} proton and deuteron data, and NuTeV $F_{2,3}$ structure function data~\cite{NuTeV:2005wsg}. We fit to both the NMC absolute cross sections in \cite{NewMuon:1996fwh} and the deuteron-proton ratio data in \cite{NewMuon:1996uwk} because the latter is obtained from roughly four times as much luminosity as the former. Since the systematic uncertainties largely cancel in the ratio in 
\cite{NewMuon:1996uwk} there remains some unknown statistical correlation with the absolute cross sections \cite{NewMuon:1996fwh} from the roughly 25$\%$ overlap in data, but we judge that the gain from independent measurements justifies us not being able to take into account this correlation. Indeed, comparing the ratio of the absolute cross sections with the ratios in \cite{NewMuon:1996uwk}, it is clear there is little correlation between the two data sets. 

\begin{figure}[t]
\begin{center}
\includegraphics[scale=0.5]{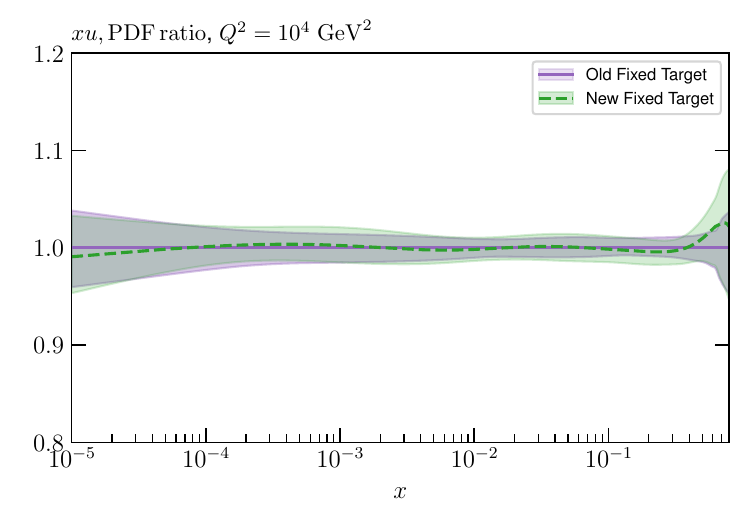}
\includegraphics[scale=0.5]{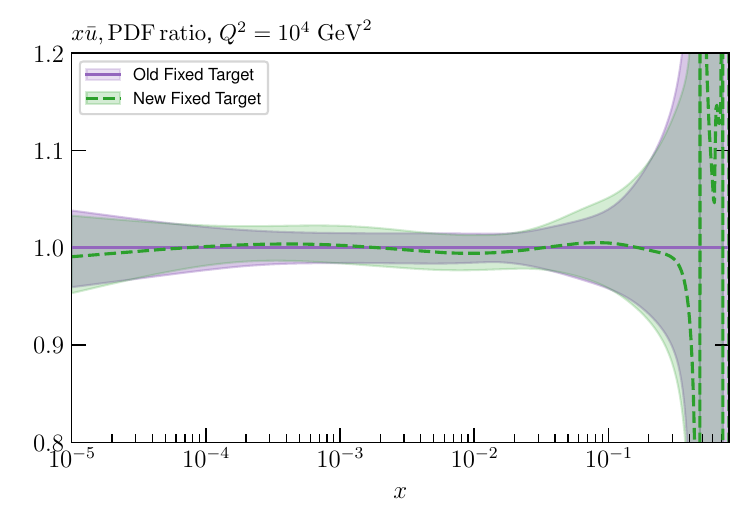}
\includegraphics[scale=0.5]{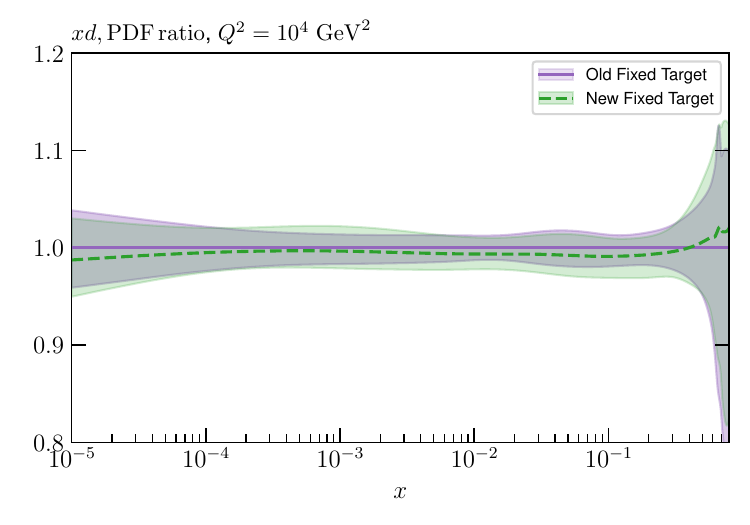}
\includegraphics[scale=0.5]{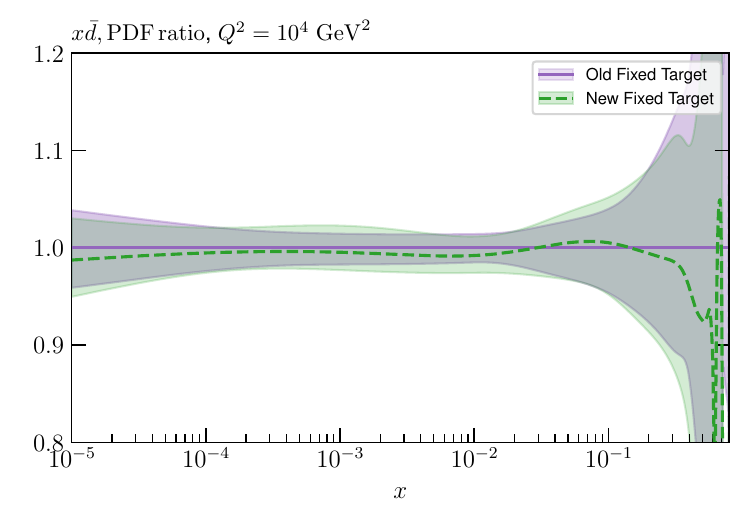}
\includegraphics[scale=0.5]{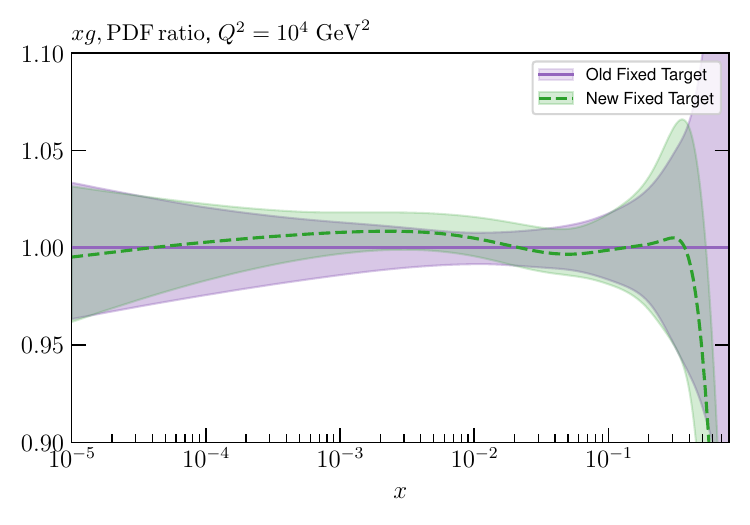}
\includegraphics[scale=0.5]{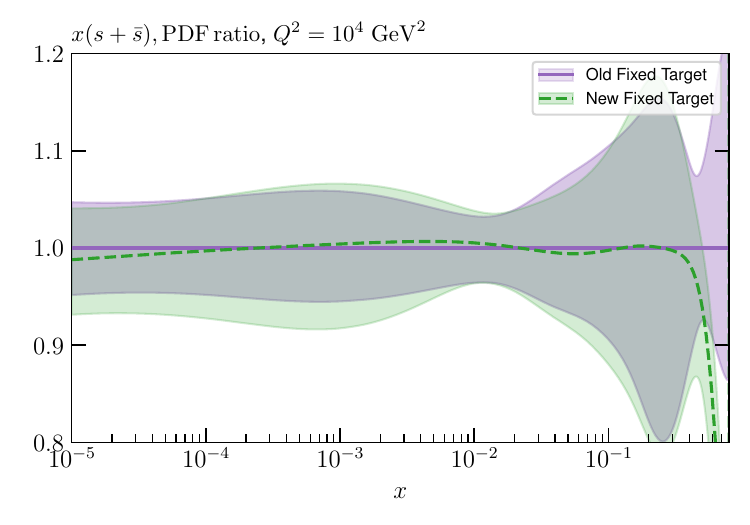}
\caption{\sf A selection of PDFs at $Q^2=10^4$ ${\rm GeV}^2$ that result from aN${}^3$LO  MSHT fits with the old and new treatment of the fixed target datasets.}
\label{fig:FTcomp_rat}
\end{center}
\end{figure}

In more detail, for both the BCDMS~\cite{BCDMS:1989qop} and NMC~\cite{NewMuon:1996fwh,NewMuon:1996uwk} data, whereas in all previous MSHT analyses the results averaged over beam energy were used, we now take the results for individual beam energies, which allows all quoted correlations in the different sources of systematic uncertainty to be accounted for. This results in roughly a doubling of the number of data points for both datasets, and an increase by 503 points in total. 
Moreover we now fit to the measured reduced cross section directly, rather then to the model--dependent extraction of $F_2$\footnote{The differences in PDFs obtained from the averaged structure functions and cross sections form different beam energies was examined in \cite{Dittmar:2005ed,Ball:2011isp} and found to be small.}.
For the E665~\cite{E665:1996mob} proton and deuteron data, seven additional sources of kinematic--dependent systematic uncertainty are now included, in addition to the overall normalization uncertainty that was previously the only one accounted for. For the NuTeV $F_{2,3}$ structure function data~\cite{NuTeV:2005wsg} an omitted normalization uncertainty of 2.1\% is now included.

The impact of these updates on the baseline fit quality is shown in Table~\ref{tab:chi2_FTcomp}. For all results which follow, for the a${\rm N}^3$LO results we use the same basic framework as in the original MSHT20a${\rm N}^3$LO fit~\cite{McGowan:2022nag}, but we now take the central values of the more up--to--date splitting functions that have become available since then, see~\cite{Falcioni:2023luc,Falcioni:2023vqq,Falcioni:2023tzp,Moch:2023tdj,Falcioni:2024xyt,Falcioni:2024qpd}. 
 We consider this to be the most sensible baseline to use, given it accounts for the significant new information these calculations provide, but we note that a complete treatment, including the remaining uncertainties on the splitting functions and an updated account of missing higher order corrections will be the subject of a future study. The impact of this new theory information on the a${\rm N}^3$LO PDFs will be outlined in~\cite{DIS2025}\footnote{We do not include the final 
calculations of the transition matrix elements \cite{Ablinger:2022wbb,Ablinger:2023ahe,Ablinger:2024xtt} which appeared following \cite{McGowan:2022nag}, as we judge that reliable inclusion of the complete matrix elements, with no inherent uncertainties, requires 
a more thorough investigation of the unknown ${\cal O}(\alpha_S^3)$ contributions to the coefficient functions $C_{i,H,q(g)}(x,Q^2/m_H^2)$ in the region $Q^2 \sim m_H^2$, and also potentially an update in the default value of the heavy quark masses $m_H$.}. We in addition include the Seaquest data, the impact of which will be discussed further in the following section, in all results which follow. In all global a${\rm N}^3$LO fit qualities, we for clarity exclude the penalty terms from the remaining unknown transition matrix element and K-factors, the impact of which is rather mild with respect to the overall trends.

We can see that for the NuTeV and E665 datasets, where the updates are relatively minor, the impact on the fit quality is very mild. For the BCDMS proton data there is little change, while for deuteron data there is some mild deterioration in the $\chi^2$ per point, but with the fit quality remaining good. The most significant change is in the NMC data, both proton and deuteron, with a marked deterioration in the fit quality observed. This is in principle perfectly possible, given the averaging over beam energies effectively treats the individual sources of systematic uncertainty as uncorrelated and hence reduces the overall constraining power of the data. We note that while the individual deuteron data is not fit by other groups, for the NNPDF4.0~\cite{NNPDF:2021njg} and CT14~\cite{Dulat:2015mca} fits the proton data (for separate beam energies) is and they find a relatively poor fit quality of $\sim 1.6$ and 1.8, respectively, in line with our updated results.

This deterioration in the fit quality for the NMC data, as well the more minor changes for the other datasets, leads to an overall deterioration in the fit quality to the FT data, even if this remains better than the fit quality to the HERA and hadron collider datasets. For the HERA data, we can see that there is a moderate deterioration in the fit quality, presumably driven by the increased constraining power of the FT data, and the known tensions between this and the HERA data. Overall, the global fit quality deteriorates by 0.04 at both orders. The trends between old and updated FT dataset treatments are rather similar between the two orders.

We next consider the impact on the PDFs that result from the fit. The change in the PDFs is shown in Fig.~\ref{fig:FTcomp_rat} and we can see that it is in most cases relatively mild. The most noticeable impact is on the gluon, with some change in shape at moderate $x$ observed, by e.g. $\sim 1\%$ in the region of relevance for the $ggH$ cross section. All changes however remaining well within the PDF uncertainty bands, which themselves are rather stable. 

\section{Impact of High $x$ ZEUS Data}\label{sec:ZEUS}

As well as the impact of the alternative description of the fixed target data sets, we also consider new data \cite{ZEUS:2020ddd} published by the ZEUS collaboration that explores this high $x$ region. This in particular uses a much finer binning than the ZEUS data entering the HERA Run II combination data~\cite{H1:2015ubc} that is used in the baseline MSHT fit (as well as other PDF analyses) and extends the kinematic region up to $x\to 1$. This enables a much more detailed analysis of this kinematic space. However, it should be noted that the total dataset presented in~\cite{ZEUS:2020ddd} contains some data in the lower $x$ region that effectively corresponds to that already included in the MSHT fit, via the combined HERA data~\cite{H1:2015ubc}. However, a procedure for removing this double counting is provided, and which we will make use of below. 

In order to access how impactful this data is on the current MSHT20 PDF sets \cite{Bailey:2020ooq} without including it in a full global fit in the first instance we use the method of Hessian re-weighting~\cite{Paukkunen:2014zia}. We note in particular that accounting for these data within the fit itself is somewhat non--trivial due to the limited event numbers in the data sample described further below.

\subsection{The Theoretical Calculation}
Following~\cite{ZEUS:2020ddd}. the predictions for the observed number of events in measured kinematic variables, $(x_{rec}, Q^2_{rec})$, are given by integrating over the full kinematic phase space:

\begin{equation}
    \nu_{j,m} = \mathcal{L} \int_{(\Delta x, \Delta Q^2)_j} [ \int A(x_{rec},Q^2_{rec} |x,Q^2)\frac{d^2\sigma(x,Q^2|PDF_m)}{d x d Q^2}d x d Q^2] \,d x_{rec} \,d Q^2_{rec}\;.
\end{equation}
where, $\mathcal{L}$ is the luminosity, $\frac{d^2\sigma(x,Q^2|PDF_k)}{dxdQ^2}$ is the differential cross-section at $(x,Q^2)$ for PDF set $m$ using the kinematic quantities defined at the Born level and $A(x_{rec},Q^2_{rec}|x,Q^2)$ transforms the Born level cross sections to observed cross sections including all relevant effects (radiative corrections, detector resolution and acceptance, selection criteria etc.). This integral is approximated to:
\begin{equation}
    \nu_{j,m} \approx \sum_i A_{ji} \lambda_{i,m}\;,
    \label{prediction}
\end{equation}
\noindent
where $\lambda_{i,m}$ is the expected number of events for the $i^{th} (\Delta x,\Delta Q^2)$ bin at the Born level for PDF set $m$, and $A_{ji}$ gives the transformation to the expectation in the measured quantities in bin $j$.

The born cross-section $\lambda_{i,m}$ is calculated by integrating across each bin $(\Delta x, \Delta Q^2)$ the complete neutral current ($\gamma$ and Z) exchange for $e^{\pm}p \rightarrow e^{\pm}p$~\cite{H1:2015ubc}.

\subsection{Evaluating the Fit Quality}

Given the fine binning of the ZEUS data, and the fact the data has been extended to $x=1$, there can be a very small number of event counts per bin. This means that the standard $\chi^2$ fitting technique is not appropriate, and instead a Poisson distribution must be used to account for the statistical fluctuations more precisely. 

From~\cite{ZEUS:2020ddd} the probability, $P(y|PDF_m)$, for a prediction based on a given PDF set $m$ to predict the data set $y$ is:
\begin{equation}
    P(y|PDF_m) = \prod_j \frac{e^{-\nu_{j,m}} \nu_{j,m}^{n_j}}{n_j !}\;,
    \label{poisson}
\end{equation}
\noindent
where the index $j$ labels the data bins $(\Delta x, \Delta Q^2)$, $\nu_{j,k}$ is the expected number of events in bin $j$ as predicted by PDF set $m$ and $n_j$ is the observed number of events. We now define our $\chi^2_{ZEUS}$ for the ZEUS data set in the following way:
\begin{equation}
    \chi^2_{ZEUS,m} = - 2  \sum_{j=1}^{N_{data}}  ln\left( \frac{e^{-\nu_{j,m}} \nu_{j}^{n_j}}{n_j !} \right) \;.
\end{equation}
We next aim to use the method of Hessian reweighting as outlined in \cite{Paukkunen:2014zia} to determine the impact of this new data on the MSHT20 PDFs at NNLO. 
To extend this formalism beyond the standard $\chi^2$ fitting approach, we define the total $\chi^2_{new}$ (i.e the sum of the $\chi^2$ for the new data and $\chi^2$ for the MSHT20 data) in a similar way to the way it has been defined in equation 3.1 of \cite{Paukkunen:2014zia}:
\begin{equation}
    \chi^2_{new} \equiv \chi^2_0 + \sum_{k=1}^{N_{eig}} (T^{\rm sym}_k)^2 w_k^2 - 2 \sum_{j=1}^{N_{data}} ln\left( \frac{e^{-\nu_{j}} \nu_{j}^{n_j}}{n_j !} \right)\;.
    \label{Total chi2}
\end{equation}
where $t^{\rm sym}_k=(T_k^++T_k^-)/2$ is the symmetrised tolerance for each eigenvector $k$.

In equation \ref{Total chi2} we have estimated the theoretical values $\nu_j$ in arbitrary $z$-space coordinates by:
\begin{equation}
   \nu_j \approx \nu_{j}[S_0] + \sum_{k=1}^{N_{eig}} D_{ik} w_k
\end{equation}
where we have defined:
\begin{equation}
    D_{ik} \equiv \frac{\nu_{i}[S_k^+] - \nu_{i}[S_k^-]}{2}
\end{equation}
$\nu_{j}[S_0]$ is the theoretical value for the $j^{th}$ bin using the base MSHT20 PDF set, and $\nu_{i}[S_k^\pm]$ is the theoretical value for the $i^{th}$ bin calculated using the plus or minus of the $k^{th}$ eigenvector of the MSHT20 data set. We can then minimize the $\chi^2_{new}$ with respect to $w_k$ to obtain the eigenvector weights, $w_{min}$.  

\subsection{Results}

\begin{figure}[t]
\begin{center}
\includegraphics[scale=0.23]{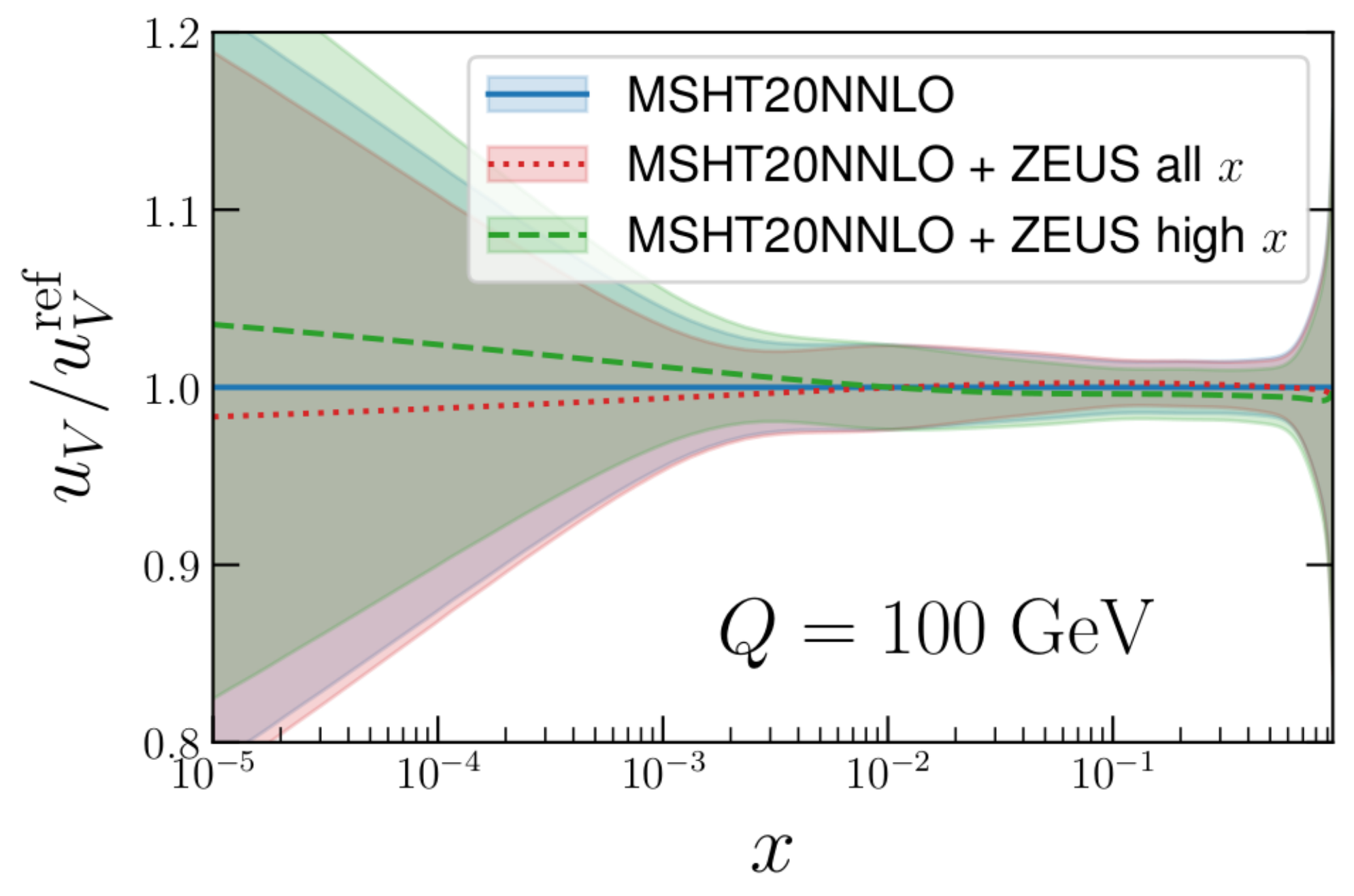}
\includegraphics[scale=0.23]{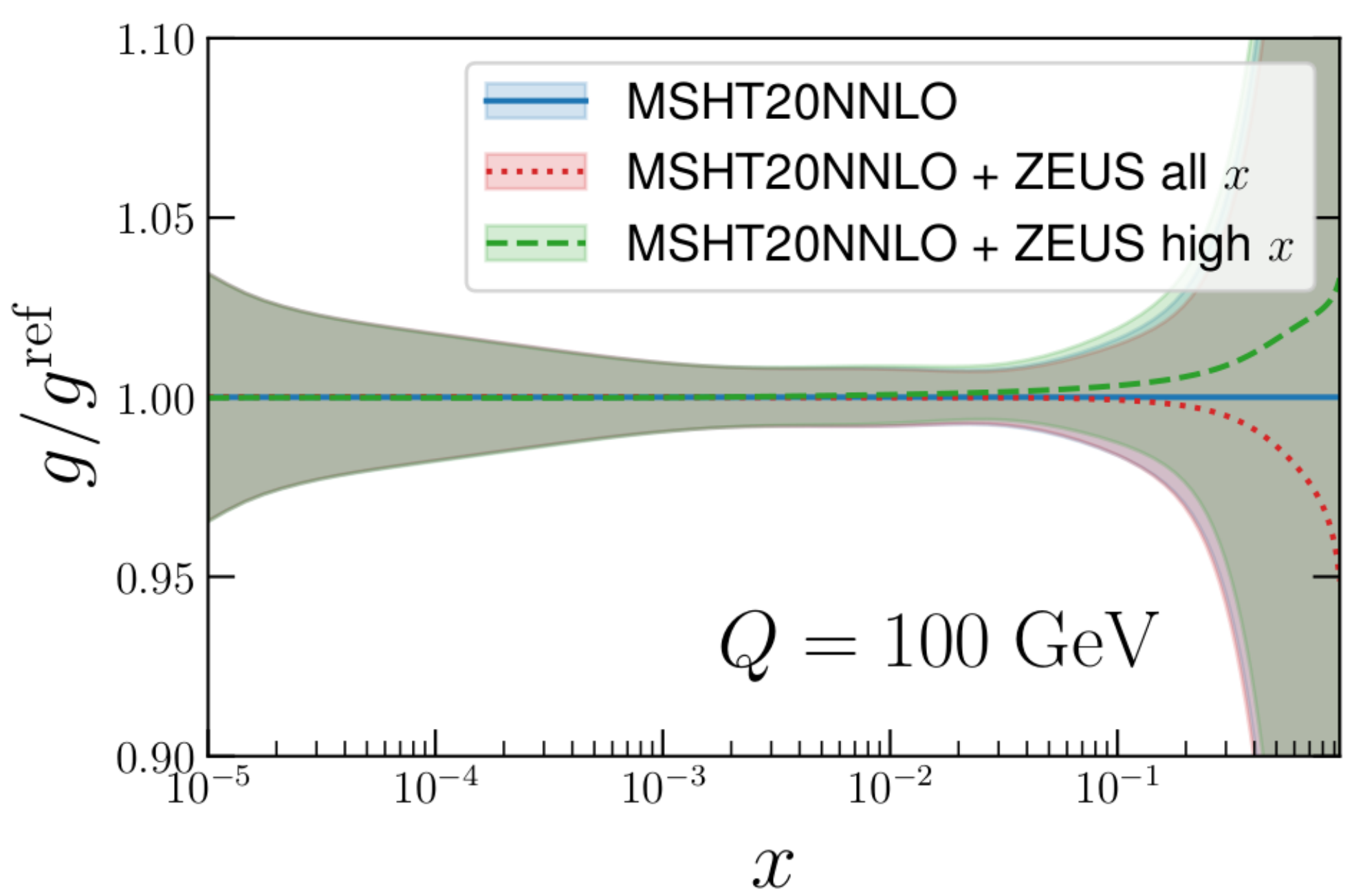}
\caption{\sf Impact of ZEUS high data~\cite{ZEUS:2020ddd} on MSHT20 baseline PDFs, via Hessian reweighting procedure. Results with the full dataset described in the reference, and with only the $x>0.35$ data included, in order to remove double counting with the HERA combined data, are shown. The gluon and $u_V$ are shown as they have the largest impact on the PDF central values.}
\label{fig:ZeusPDF}
\end{center}
\end{figure}

Applying the procedure described above we find that in general the impact of these data is rather limited. In Fig.~\ref{fig:ZeusPDF} we show the impact on the absolute value of the PDFs, for both the full ZEUS dataset and with the lower $x$ removed, in order to avoid double counting. The latter case is therefore the more appropriate, with the former given for demonstration. In all cases, we include both the positron and electron data together.

In Fig.~\ref{fig:ZeusPDF} we show the impact on the absolute values of the PDF, and their uncertainties, for the up valence and gluon, which show the largest changes. We can see that this is very mild, in particular once the high $x$ data alone are included which removes the double counting with the existing HERA combined data in the fit 

\begin{figure}
\begin{center}
\includegraphics[scale=0.23]{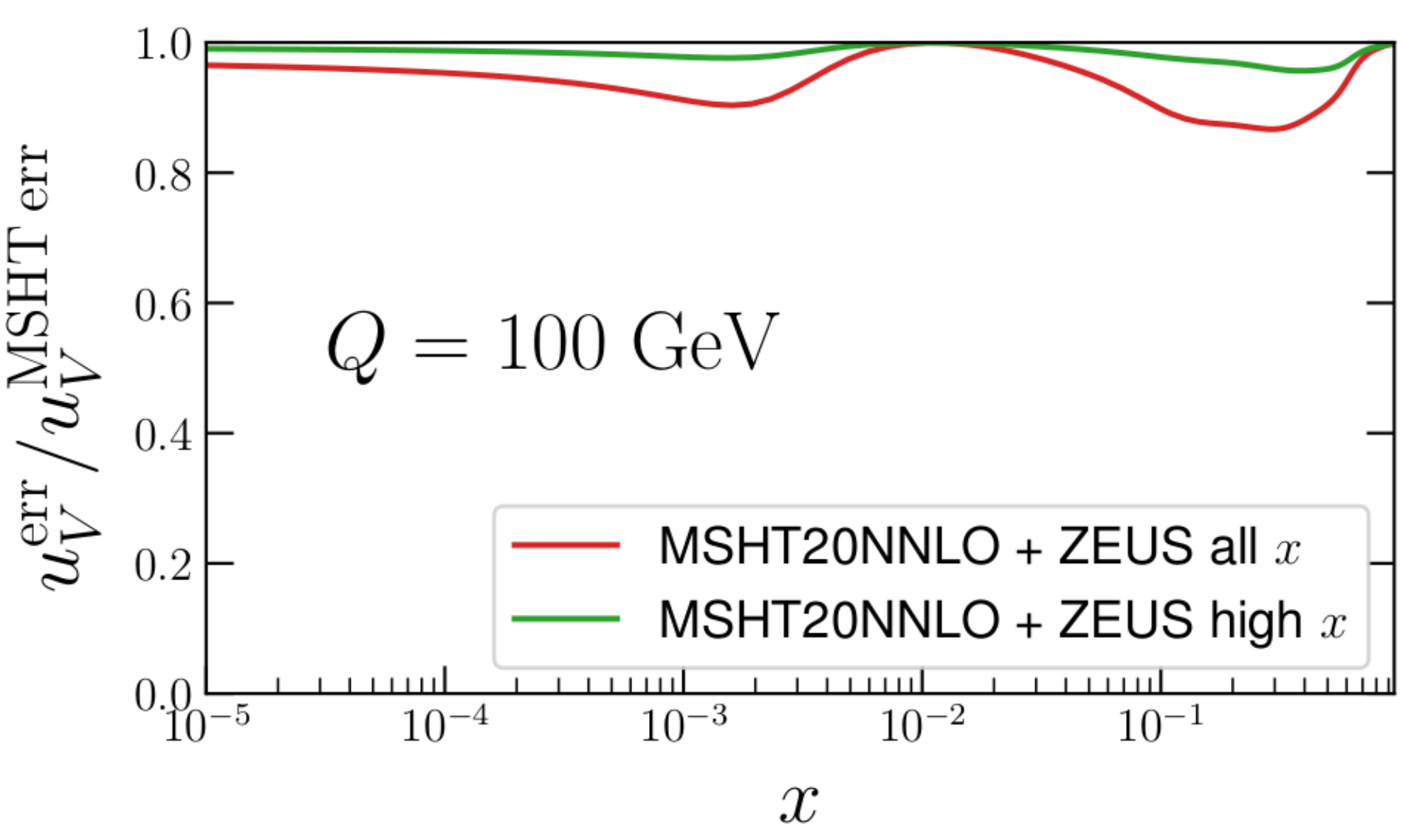}
\includegraphics[scale=0.23]{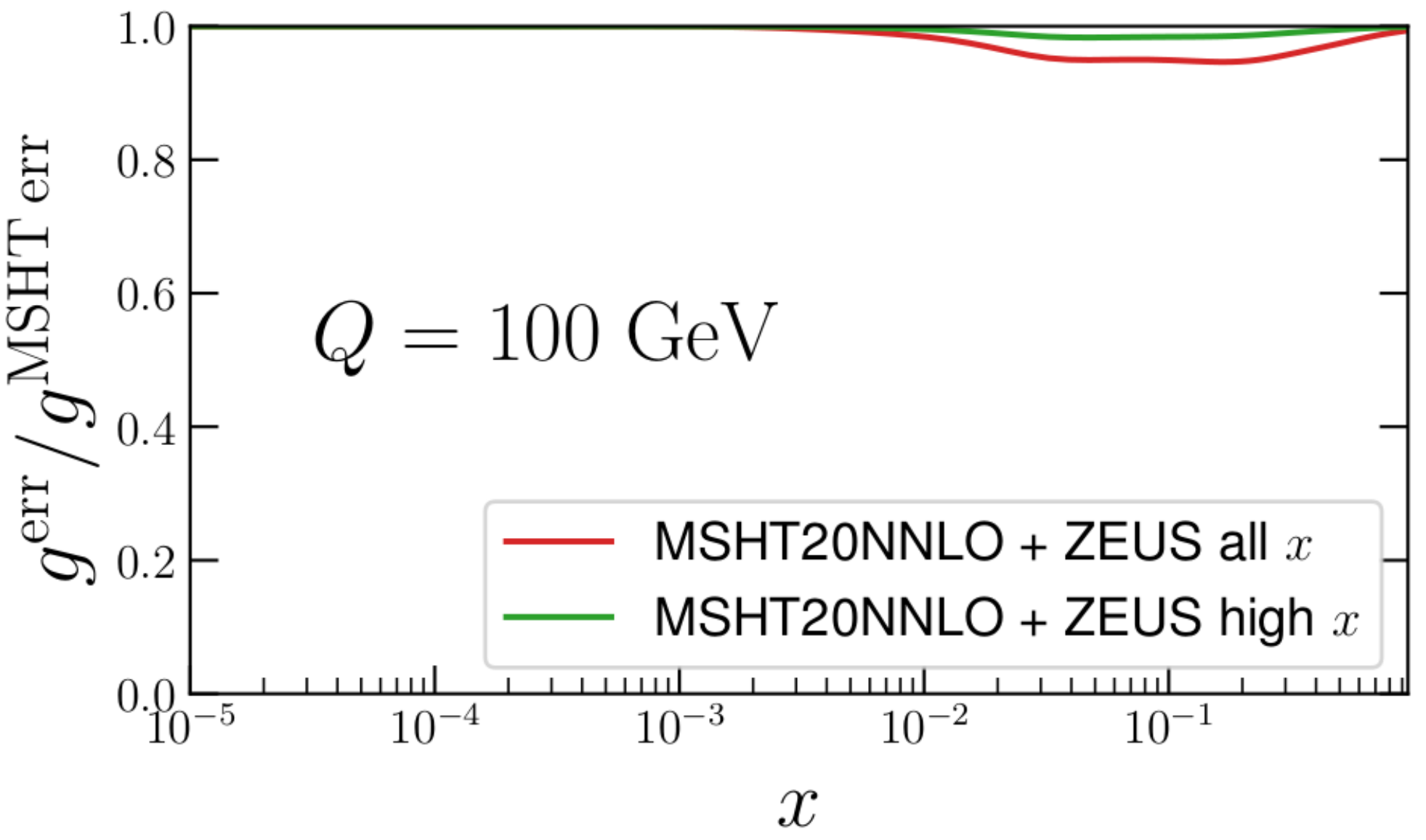}
\caption{\sf As in Fig.~\ref{fig:ZeusPDF} but for ratio of the PDF uncertainty to the MSHT20 baseline.}
\label{fig:ZEUSunc}
\end{center}
\end{figure}

The impact on the PDF uncertainties is shown in Fig.~\ref{fig:ZEUSunc} for the up valence and gluon, again as these show the largest effect. This is again rather mild, reducing the uncertainty by at most a few percent once the high $x$ data alone are included.
When the lower $x$ data are removed, we note that there is still some very mild impact at lower $x$ on the up valence, presumably due to the sum rules.

In terms of the penalty term, defined in~\cite{Paukkunen:2014zia} as:
\begin{equation}
P \equiv \sum_{k=1}^{N_{eig}} \left[ \left( \frac{t_k^+ + t_k^-}{2}\right) w_k^{min} \right]^2
\label{P def}
\end{equation}
we find $P\approx 1.4$ for the full dataset, and similarly if slightly smaller when the high $x$ contribution alone is included. As this is rather smaller than the $\Delta \chi^2$ corresponding to the average MSHT20 tolerance, this implies that there will only be a relatively small impact on the MSHT20 PDF sets by adding this data, and any modification would fall well within the errors of the MSHT20 PDF set. Indeed, we have verified this result explicitly above.

\begin{figure}[t]
\begin{center}
\includegraphics[scale=0.23]{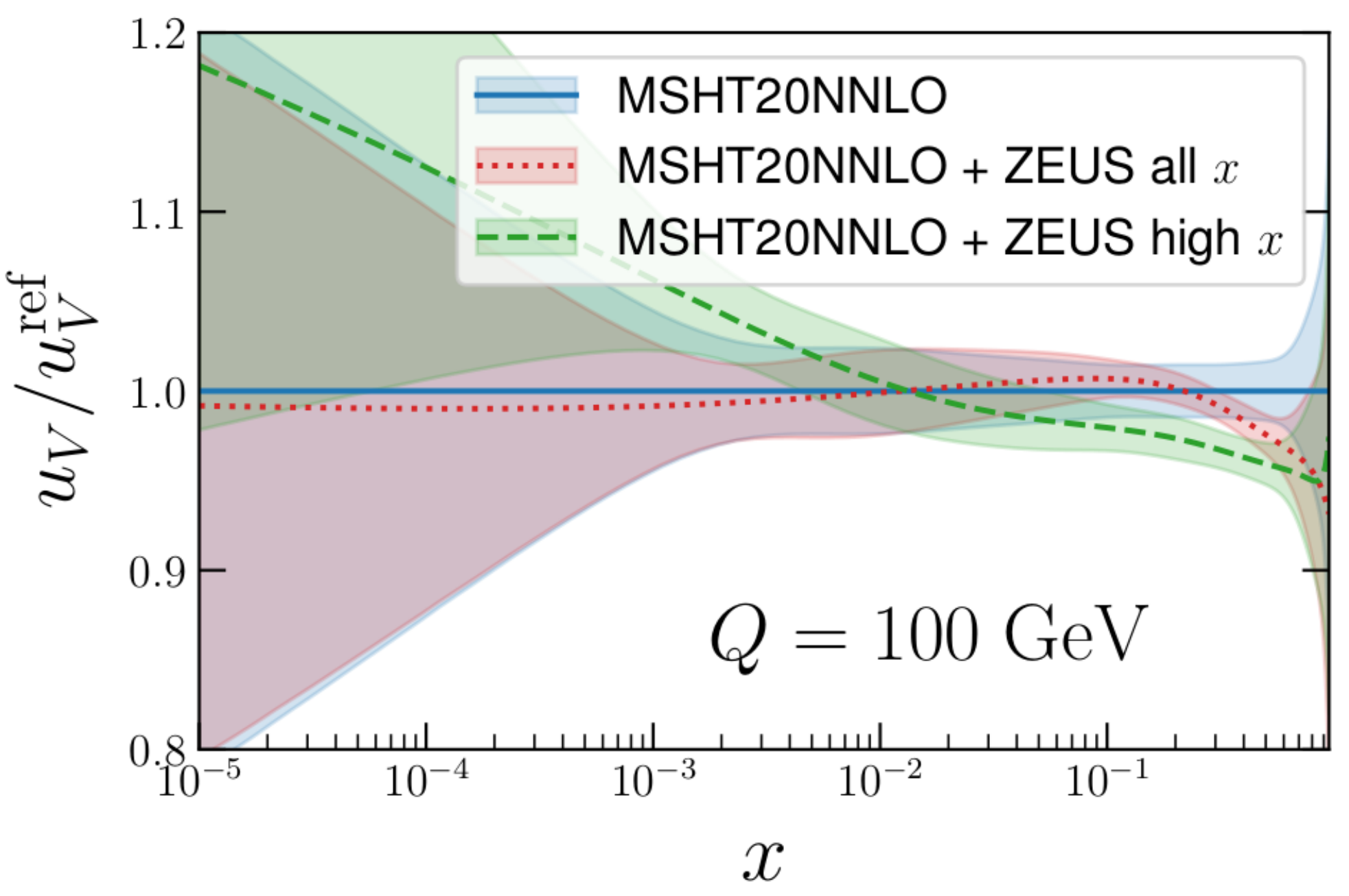}
\includegraphics[scale=0.23]{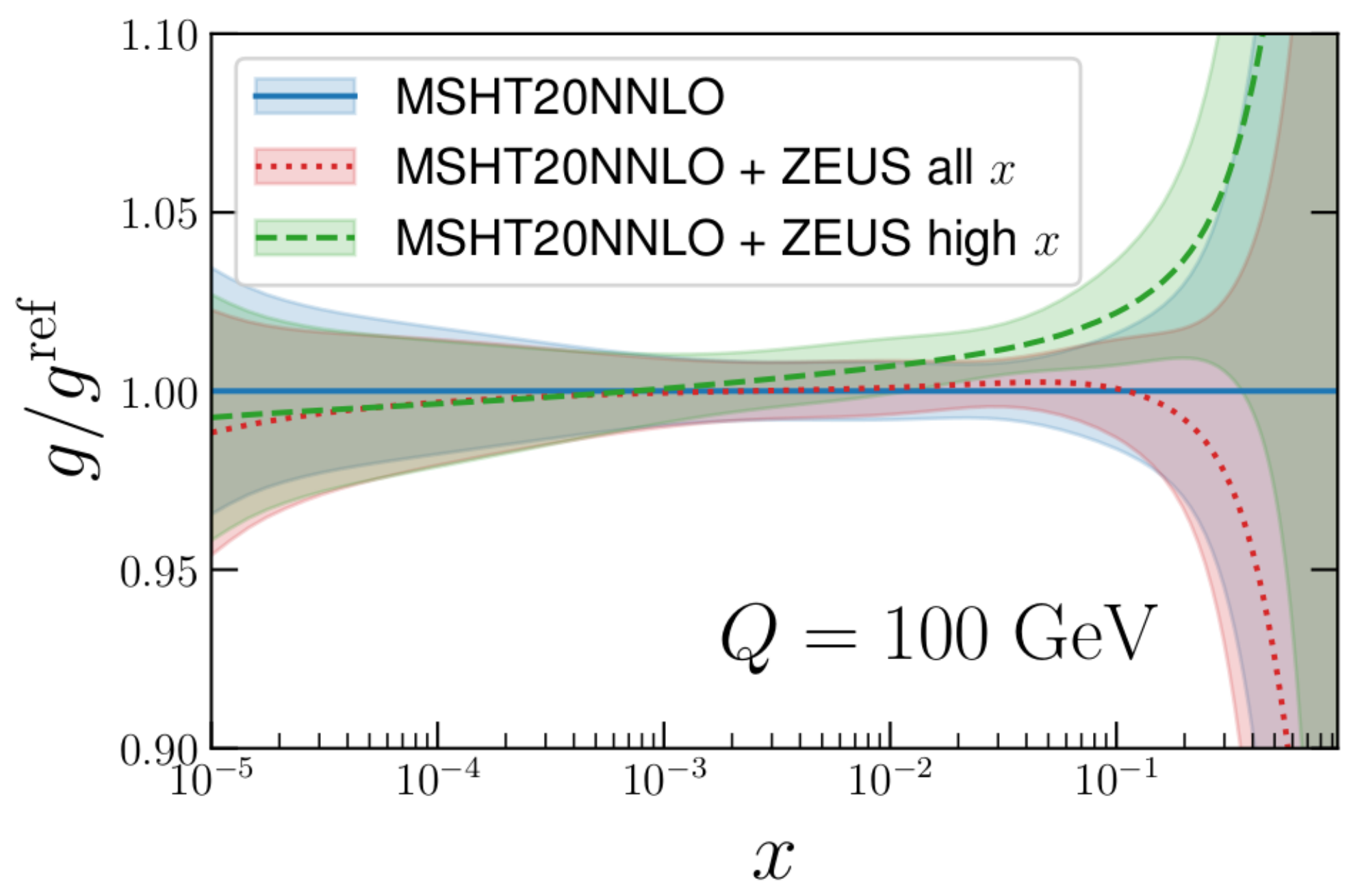}
\caption{\sf As in Fig.~\ref{fig:ZeusPDF} but with the tolerance factors incorrectly set to 1 in the profiling.}
\label{fig:ZeusPDF_T1}
\end{center}
\end{figure}

\begin{figure}
\begin{center}
\includegraphics[scale=0.23]{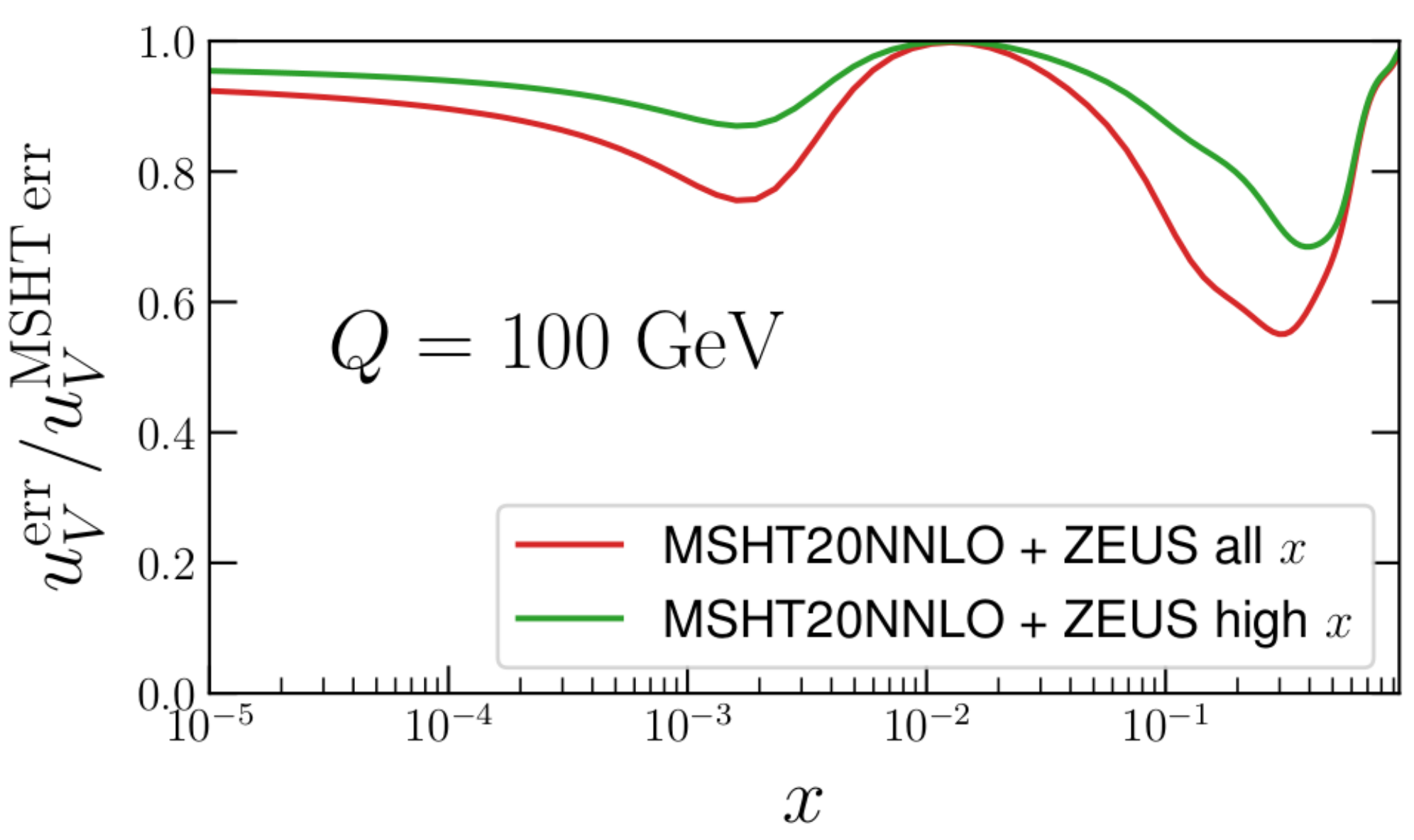}
\includegraphics[scale=0.23]{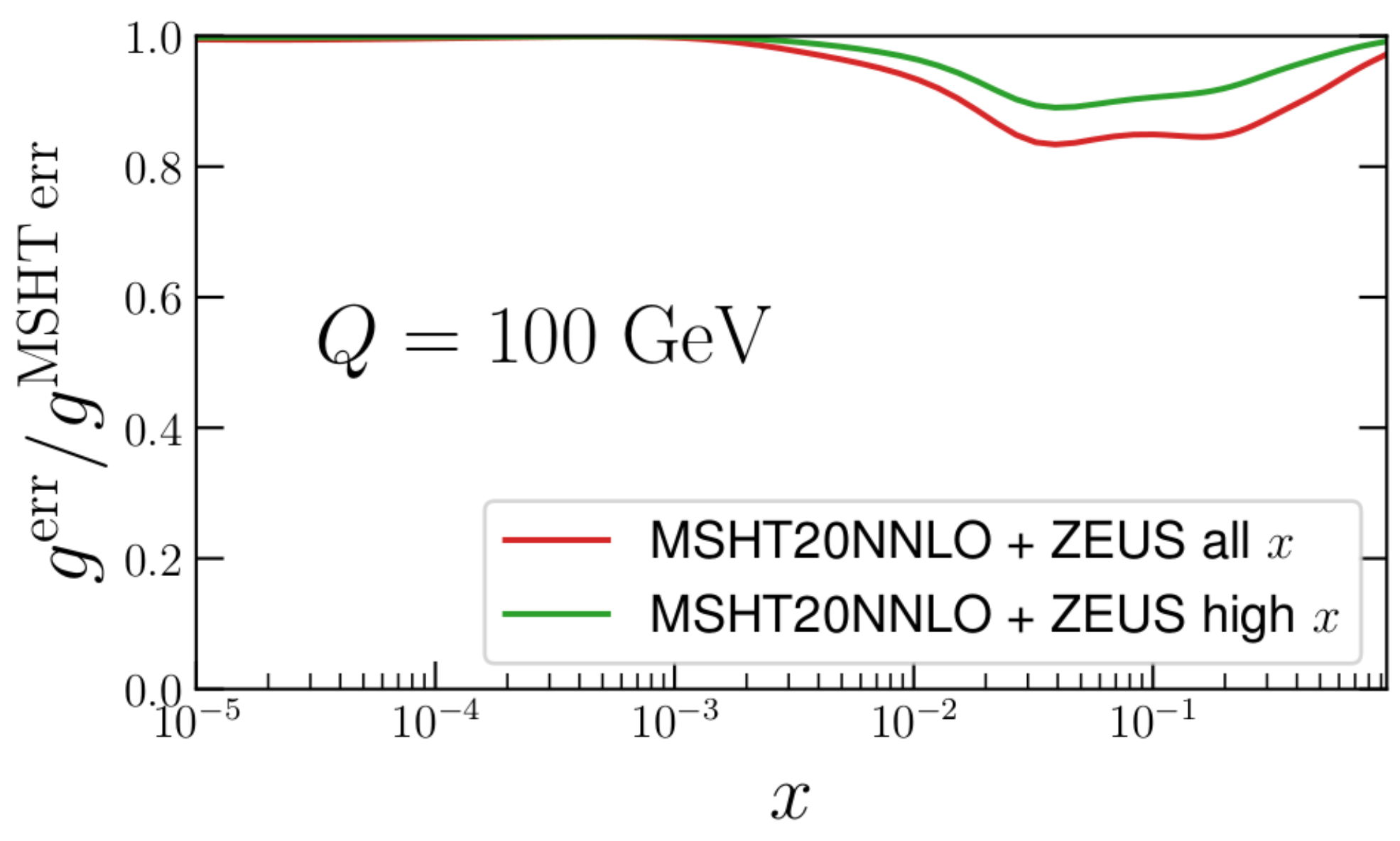}
\caption{\sf As in Fig.~\ref{fig:ZEUSunc} but with the tolerance factors incorrectly set to 1 in the profiling.}
\label{fig:ZEUSunc_T1}
\end{center}
\end{figure}

Finally, we note in particular here that in this profiling we have included the symmetrised dynamical tolerances of the MSHT20 NNLO PDFs as in~\eqref{Total chi2}, as one must to correctly reproduce the input PDF uncertainties of the profiled PDF set \cite{Paukkunen:2014zia,Schmidt:2018hvu,Hou:2019gfw,AbdulKhalek:2019mps}. Not including such tolerance factors is inconsistent with the fact that the baseline MSHT fit applies a dynamic tolerance procedure, and in particular that the prior penalty that applies in the profiling procedure assumes this. If instead we were to neglect the tolerance, as is often erroneously done in experimental analyses, we would obtain the plots in Figures~\ref{fig:ZeusPDF_T1},~\ref{fig:ZEUSunc_T1}. With the tolerance neglected (i.e. $T=1$ taken) much larger impacts on both the central values of the PDFs and the PDF uncertainties are observed. Pulls on the PDFs are now larger than the uncertainty band across several regions of $x$ for the gluon and up valence in Figure~\ref{fig:ZeusPDF_T1}, compared to much smaller pulls in Figure~\ref{fig:ZeusPDF}. Interestingly, these can be somewhat larger in the high $x$ case, despite its smaller impact on the uncertainty, presumably due to some difference in pull between the lower and higher $x$ data. In addition one would falsely conclude that the constraints placed on the PDFs of this ZEUS data are several times larger on the up valence and gluon PDFs comparing Figure~\ref{fig:ZEUSunc_T1} with the correct Figure~\ref{fig:ZEUSunc}. These comparisons should therefore serve as a warning against such an approach. We note in particular that we can calculate the penalty $P$ in this case, and find it is about $110$ for the high $x$ data, indicating a rather severe deterioration in the description of the remaining data in the fit. 

Given the non-standard procedure needed to include these data within the global fit, and the complications regarding double-counting, we do not plan to automatically include them within our choice of data sets. In the future, the impact of these data will likely be lower still, due to additional constraints from new data. We note, moreover, that this dataset will certainly be superseded by EIC data \cite{Armesto:2023hnw}.

\section{Impact of Seaquest Data}\label{sec:SQ}

A complementary probe of the high $x$ structure of the proton can be provided through fixed-target Drell-Yan experiments. In particular there has been a long-standing interest on the antiquark flavour asymmetry. This was originally assumed to be zero, i.e. $\bar{d}(x) = \bar{u}(x)$ due to flavour independence and the near identical phase space for perturbative gluon radiation to generate up and down antiquarks through $g \rightarrow q \bar{q}$. However, measurements at NMC\cite{PhysRevLett.66.2712,PhysRevD.50.R1} of the Gottfried Sum Rule \cite{PhysRevLett.18.1174} relating to the integral of the difference of charged lepton-nucleon DIS on protons to neutrons revealed this not to be the case, with the integral $\int_0^1(\bar{u}(x)-\bar{d}(x))dx < 0$ (extrapolating beyond the measured $x$ range). This implied that $\bar{d} > \bar{u}$ over at least a portion of the $x$ range, but provided no information on its $x$ dependence. There are now a variety of theoretical models that predict a non-zero flavour asymmetry of the light-quark sea of the proton, i.e. $\bar{d} \neq \bar{u}$, supported also by  lattice determinations \cite{Lin_2015}. These typically result in $\bar{d} > \bar{u}$ at intermediate to large $x$ \cite{Alberg_2019}. Various reviews of the experimental and theoretical status are available \cite{Garvey_2001,Chang_2014,Geesaman_2019}.

\begin{figure}
\begin{center}
\includegraphics[height=5.5cm,width=7.9cm]{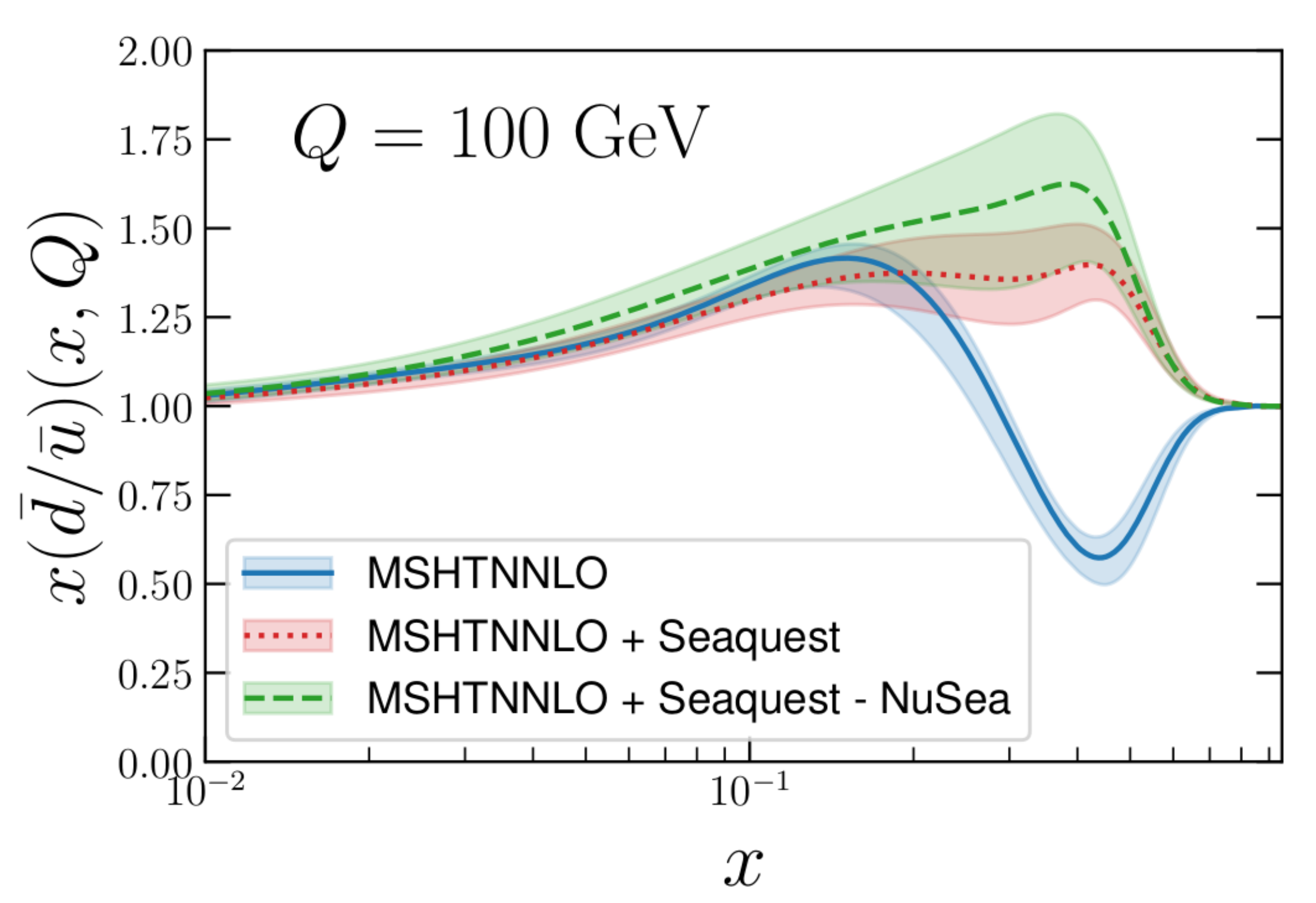}
\includegraphics[height=5.5cm,width=7.9cm]{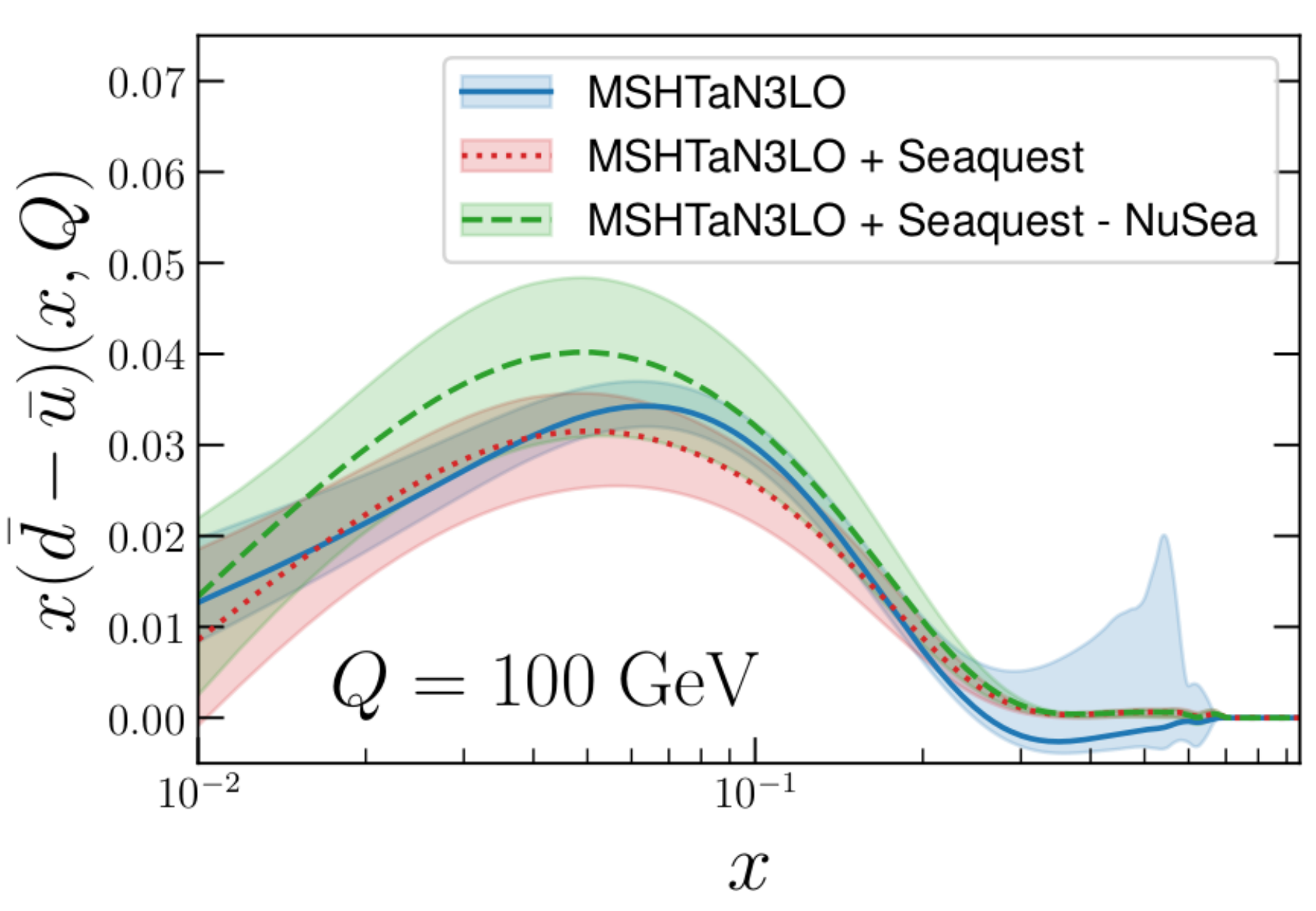}
\caption{\sf The impact of adding the Seaquest data, and in turn also removing the NuSea Drell-Yan ratio data, on the quark sea flavour asymmetry at high $x$. (Left) The effect on the NNLO PDFs illustrated by the $\bar{d}/\bar{u}$ ratio, and (right) the effect on the aN${}^3$LO PDFs illustrated by the $\bar{d}-\bar{u}$ difference, both at $Q=100{\rm GeV}$. For the latter, in the absence of the Seaquest data the $\bar{d}$ PDF goes negative at very large $x$\cite{McGowan:2022nag}, hence the difference rather than ratio of $\bar{d}$ and $\bar{u}$ is shown for ease of interpretability.}
\label{fig:Seaquest_PDFs}
\end{center}
\end{figure}

Fixed target Drell-Yan measurements are ideally suited to probe this antiquark sea asymmetry in the large $x$ region. Fixed target neutral-current Drell-Yan on protons is dominated by up valence in the beam and hence anti-up in the target. Instead performing interactions with a neutron and assuming isospin symmetry then the cross section is dominated by down valence and down antiquarks in the beam and target respectively. Therefore the ratio of the cross section on protons to that on deuterons provides direct sensitivity to $\bar{d}/\bar{u}$. As a result, the NuSea/E866 fixed target Drell-Yan experiment at Fermilab subsequently performed such measurements~\cite{Towell_2001} on dimuon data; these are included in the MSHT20 PDFs~\cite{Bailey:2020ooq,McGowan:2022nag}, and by default in our studies in the other sections of this paper. Whilst a preference for $\bar{d}(x)>\bar{u}(x)$ was observed for $x$ up to around 0.2, above this hints of the (unexpected) opposite trend $\bar{u}(x)>\bar{d}(x)$ were observed, albeit with limited statistics. These data then result in $\bar{d}/\bar{u}(x) < 1$ in MSHT for $0.25 \lesssim x \lesssim 0.75$, as seen in the ``MSHTNNLO'' baseline in the left plot of Figure~\ref{fig:Seaquest_PDFs}.

In order to investigate this the Seaquest/E906\cite{Dove_2021,Dove_2023} experiment at Fermilab repeated this measurement, though with reduced beam energies in order to probe the higher $x$ region. In their own~\cite{Dove_2021,Dove_2023} and subsequent studies by CT~\cite{Guzzi:2021fre,Hou_2022,Hou_2023}, JAM~\cite{Cocuzza_2021}, CJ~\cite{Park:2021kgf,Accardi_2023}, and ABMP~\cite{Alekhin:2023uqx}, these data have observed a similar behaviour to NuSea over the moderate $x$ region of $\bar{d}(x)>\bar{u}(x)$, however there is a slight tension at larger $x$ with the Seaquest data favouring $\bar{d}(x)>\bar{u}(x)$ continuing at large $x$. Here we investigate this in MSHT (having previously shown preliminary results) at both NNLO and approximate N3LO. 

The impact of including the Seaquest data\footnote{We take the data provided in \cite{Dove_2021} Table I, including the covariance matrix provided in equation~(9), a 2\% relative beam normalisation uncertainty, and the acceptance corrections applied as in equation~(10) provided in Extended Data Table 3.} on the MSHT PDFs is shown in Figure~\ref{fig:Seaquest_PDFs}. We note that here our baseline is the same as elsewhere in this paper, and is therefore updated with respect to the public MSHT20 PDFs\cite{Bailey:2020ooq}, as described above. At NNLO, in the left of the figure, the aforementioned behaviour of the $(\bar{d}/\bar{u})(x)$ asymmetry inherited from the NuSea data is observed, including $(\bar{d}(x)/\bar{u})(x\gtrsim 0.25)<1$. Adding the Seaquest data, this effect immediately disappears, with $(\bar{d}/\bar{u})(x) \geq 1$ over the whole of $x$. Moreover, the PDF error bands do not overlap with the previous determination, indicating a tension between the NuSea and Seaquest data at large $x$ (this is also noted in\cite{Jing:2023isu}). As a result, further removing the NuSea data\footnote{Here we refer to the E866 Drell-Yan ratio data as ``NuSea'', it is this data alone we remove, we do not remove the related E866 Drell-Yan absolute data~\cite{webb2003measurementcontinuumdimuonproduction}.} from the fit whilst retaining the Seaquest data raises the $\bar{d}/\bar{u}$ somewhat further, albeit with its uncertainty band still covering the PDF extracted with both NuSea and Seaquest included. On the other hand, the PDFs in all cases are consistent below $x \approx 0.2$, showing the agreement of the measurements outside the large $x$ region.

Repeating this analysis at a${\rm N}^3$LO we obtain Figure~\ref{fig:Seaquest_PDFs}~(right). The default a${\rm N}^3$LO PDFs have a slightly negative $\bar{d}$ at high $x \gtrsim 0.4$ (as noted in~\cite{McGowan:2022nag} and reflected in the unusual shape error band in the ``MSHTaN3LO'' baseline above $x \gtrsim 0,4$ in the figure), therefore we show here the difference of the down and up antiquarks $(\bar{d}-\bar{u})(x)$ for ease of presentation. The trend observed upon addition of the Seaquest data is the same, with $\bar{d}-\bar{u}$ raised such that it is now positive over the whole $x$ range and also prevents the $\bar{d}$ from becoming negative (not shown). Again,  removing the NuSea data reinforces this behaviour further. We note that at a${\rm N}^3$LO the PDF uncertainty for the baseline does cover the result upon addition of Seaquest at high $x$, unlike at NNLO.

The tension of the NuSea and Seaquest data at large $x$ can also be seen at the level of the fit qualities, provided in Table~\ref{tab:chi2_Seaquest}. At both NNLO and a${\rm N}^3$LO the NuSea data $\chi^2$ doubles upon addition of the Seaquest data, which reflects also the smaller uncertainties of the latter data. At the same time the $\chi^2$ of the rest of the data in the global fit increases - by 14 units at NNLO and by 24 units at a${\rm N}^3$LO. In both cases (NNLO and a${\rm N}^3$LO) the patterns are the same, the increases are focused mainly on the fixed target DIS data with the BCDMS and NMC worsening by around 10 units. The D0 $W$ asymmetry data, which is sensitive to high $x$ quark PDFs, also worsens, on the other hand some of the LHC Drell-Yan data improve slightly. Upon removing the NuSea ratio data this tension with other data in the fit reduces. However, the Seaquest $\chi^2$ worsens somewhat, perhaps related to tension with other data in the region at moderate $x$ where the NuSea ratio and Seaquest data agree.

\begin{table}
\begin{center}
   \scriptsize
   \centering
    \renewcommand{\arraystretch}{1.4}
 \begin{tabular}{|>{\centering\arraybackslash}m{1.65cm}|>{\centering\arraybackslash}m{0.6cm}|>{\centering\arraybackslash}m{1.25cm}|>{\centering\arraybackslash}m{1.45cm}|>{\centering\arraybackslash}m{1.8cm}|>{\centering\arraybackslash}m{1.25cm}|>{\centering\arraybackslash}m{1.45cm}|>{\centering\arraybackslash}m{1.8cm}|}
\hline 
& & \multicolumn{3}{>{\centering\arraybackslash}m{4.5cm}|}{NNLO} & \multicolumn{3}{>{\centering\arraybackslash}m{4.5cm}|}{aN${}^3$LO} \\ \hline
$\chi^2$ & $N_{\rm pts}$ & baseline & +Seaquest & +Seaquest - NuSea ratio & baseline & +Seaquest & +Seaquest -NuSea ratio \\ \hline
NuSea ratio & 15 & 9.7 & 20.2 & - & 8.1 & 18.3 & - \\ \hline
Seaquest & 6 & - & 6.9 & 7.6 & - & 9.3 & 13.6 \\ \hline
Total -NuSea ratio -Seaquest & 4519 & 6296.3 & 6310.5 & 6299.5 & 6159.9 & 6187.0 & 6173.4\\ \hline
Total & 4534 & 6306.1 & 6337.7 & 6307.1 & 6168.0 & 6214.7 & 6187.0 \\ \hline 
 \end{tabular}
\end{center}
 \caption{\sf $\chi^2$ values for MSHT fits at NNLO and aN${}^3$LO with the Seaquest data added and in turn also the NuSea Drell-Yan ratio data removed, relative to the baseline case which includes only the NuSea ratio data.
}
 \label{tab:chi2_Seaquest}
\end{table}

\section{Power Corrections and Cut Dependence}\label{sec:TMCS}

In this section we consider the impact of power corrections in $1/Q^2$ for the DIS data entering the MSHT fit. One known source of these derives from TMCs, which are due to the kinematic effect of the non--zero hadron mass. These can be calculated in terms of the standard leading twist results~\cite{Georgi:1976ve}, see \cite{Schienbein:2007gr} for a summary. We include the dominant such correction, which affects $F_2$ following the approximate expression from this reference, namely
\be
F_2^{\rm TMC}(x,Q^2) \approx \frac{x^2}{\xi^2 r^3} F_2^{(0)}(\xi,Q^2)\left[1+ \frac{6 x\mu\xi}{r}(1-\xi)^2\right]\;,
\ee
where $\xi$ is the usual Nachtmann variable. We have investigated the effect of including corrections to the longitudinal structure function, $F_L$, and find this to have a negligible impact on the fit.

In addition to these known corrections, there may arise contributions from less well understood higher--twist (HT) corrections from correlations of the partons inside the hadron. Following numerous studies on these~\cite{Virchaux:1991jc,Martin:2003sk,Blumlein:2006be,Alekhin:2012ig,Thorne:2014toa,Cerutti:2025yji} we parameterise these in an $x$--dependent manner, multiplicatively, via
\be\label{eq:HT}
F_2^{\rm HT}(x,Q^2)=F_2^{\rm LT}(x,Q^2)\left(1+ \frac{D_i(x)}{Q^2}\right)\;,
\ee
where the $D_i$ correspond to the HT corrections to $F_2$, and again we can safely neglect corrections to $F_L$ and/or $F_3$ for the purposes of our study. To be precise, we take the same approach as in~\cite{Martin:2003sk} and treat these $D_i$ as a set of 12 free parameters in different $x$ bins, ranging from $x < 0.005$ to $0.7 < x < 0.8$\footnote{We omit the correction in the 13th, highest $0.8< x <0.9$, bin as this is only sensitive to handful of datapoints, with the lower $W^2$ cut, and none with the higher cut, and hence leads to rather unstable results for the corresponding HT correction, with little impact on the fit.}. These parameters can then be allowed to vary freely in the fit, in order to determine the corresponding (effective) HT correction. As we will see below, the interpretation of these  is not always clearly in terms of purely HT corrections, but is rather tied up with missing higher order QCD corrections that can be absorbed in the fit coefficients. However, for the sake of brevity we will often in what follows refer to these as HT corrections, even if their origin is not necessarily higher twist.

\begin{table}
\begin{center}
  \scriptsize
  \centering
   \renewcommand{\arraystretch}{1.4}
\begin{tabular}{Xrccc}\hline 
&Baseline cuts, no TMC &  no TMC, lower $W^2$ cut & w. TMC, lower $W^2$ cut
\\ \hline
BCDMS $p$& 360.5 (1.10) & 423.5 (1.21) & 417.9 (1.19) \\
BCDMS $d$& 251.6 (1.02) & 313.8 (1.24)  & 297.0 (1.17) \\
NMC $p$ &383.9 (1.57) & 416.8 (1.62) & 416.4 (1.61) \\
NMC $d$&326.3 (1.34) & 361.1 (1.40) & 361.0 (1.40)\\
SLAC $p$&31.3 (0.85)& 128.2 (1.07) & 107.1 (0.89) \\
SLAC $d$& 22.2 (0.58)& 78.3 (0.67) & 52.7 (0.45)\\
E665 $p$&  75.0 (1.41) &76.2 (1.44) & 77.0 (1.45)\\
E665 $d$& 72.0 (1.36)  & 69.6 (1.31) & 70.7 (1.33)\\
NuTeV $F_2$ &  35.6 (0.67)& 38.3 (0.70) & 37.1 (0.67)\\
NuTeV $F_3$ & 33.3 (0.79) & 33.9 (0.81) & 31.1 (0.74)\\
NMC $n/p$& 144.4 (0.98) & 176.0 (1.00) & 172.2 (0.98)\\
{\bf Fixed Target}&{\bf 2201.1 (1.11)}  & {\bf 2590.7 (1.15)} & {\bf 2511.1 (1.12)} 
\\ \hline
{\bf HERA}& {\bf 1626.9 (1.29)}& {\bf 1648.0 (1.30)}  & {\bf 1616.4 (1.28)} \\
{\bf Hadron Collider}& {\bf 2386.8 (1.33)}& {\bf 2407.7 (1.35)}& {\bf 2399.1 (1.34)}
\\ \hline \hline 
{\bf Global }  & \bf{6214.7 (1.23)} & \bf{6646.4 (1.26)}& \bf{6526.6 (1.23)}\\
\hline
\end{tabular}
\end{center}
\caption{\sf $\chi^2$ values for MSHT fits,  with target mass corrections included or excluded, and with a lower (default) $W^2$ cut of 5 (15) ${\rm GeV}^2$, as indicated. The absolute value is given, along with the $\chi^2$  per point in brackets, for the individual fixed target datasets, as well as for the results for the global dataset, and subsets of it. The lower (higher) $W^2$ cut corresponds to 2239 (1983) fixed target data points.}
\label{tab:chi2_TMCcomp}
\end{table}

For all results which follow we apply the updated treatment of the FT datasets discussed in the previous section. In the default MSHT fits, a cut of $Q^2 > 2\,{\rm GeV^2}$ and $W^2>15 \,{\rm GeV^2}$ is imposed (with higher $W^2> 25\, {\rm GeV^2}$ cut for $F_3$ determinations) in order to limit the impact of such power corrections, which are not modelled in the fit. If we do model TMCs and HT corrections, we may be able to place a less conservative cuts, and hence we also explore the impact of imposing a lower cut of  $W^2> 5\, {\rm GeV^2}$ (keeping the $F_3$ cut fixed). 

\subsection{Fit Quality and PDF Impact}\label{sec:fitqual}

We start by considering the impact of TMCs alone. The impact on the fit quality is shown in Table~\ref{tab:chi2_TMCcomp}. The impact of including these for the baseline cuts (not shown) is very mild, with the fit quality improving by a handful of points. This is as we may expect, given the $W^2$ cut will largely remove sensitivity to such corrections. Once the $W^2$ cut is lowered, we can see that absent TMCs the fit quality deteriorates significantly, by about $1-2\sigma$ globally and for the FT data. However, once these are included there is a dramatic improvement in the fit quality, and indeed the $\chi^2$ per point globally and for the various subsets is very similar to the baseline case, i.e. with the higher $W^2$ cut and absent TMCs. Therefore, at the level of the fit quality  the inclusion of TMCs alone is sufficient to give a comparable result to the baseline; although as we will see (and indeed has been observed elsewhere~\cite{Virchaux:1991jc,Blumlein:2006be,Alekhin:2012ig}) there is a  preference for non--zero HTs on top of this.

\begin{figure}[t]
\begin{center}
\includegraphics[scale=0.5]{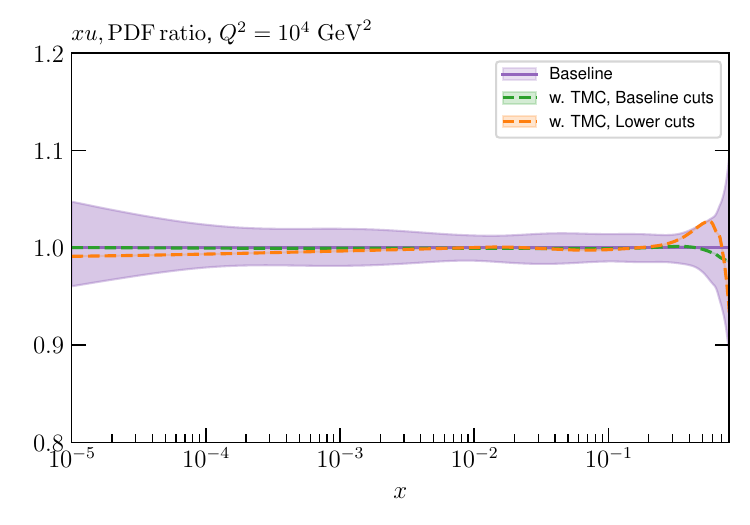}
\includegraphics[scale=0.5]{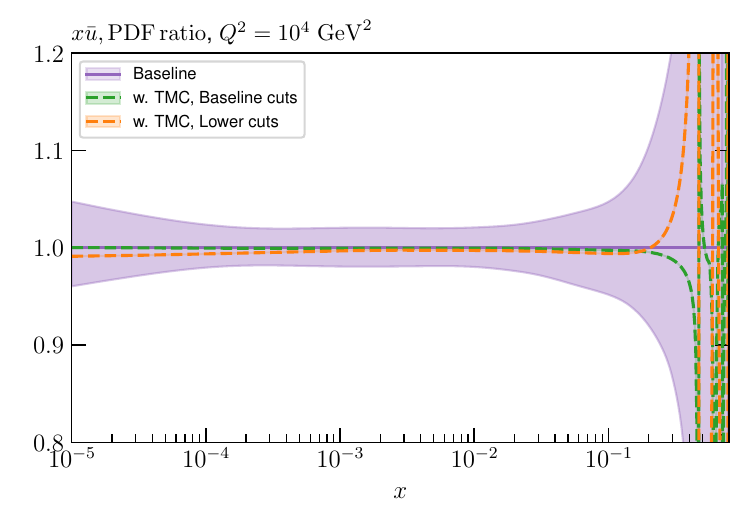}
\includegraphics[scale=0.5]{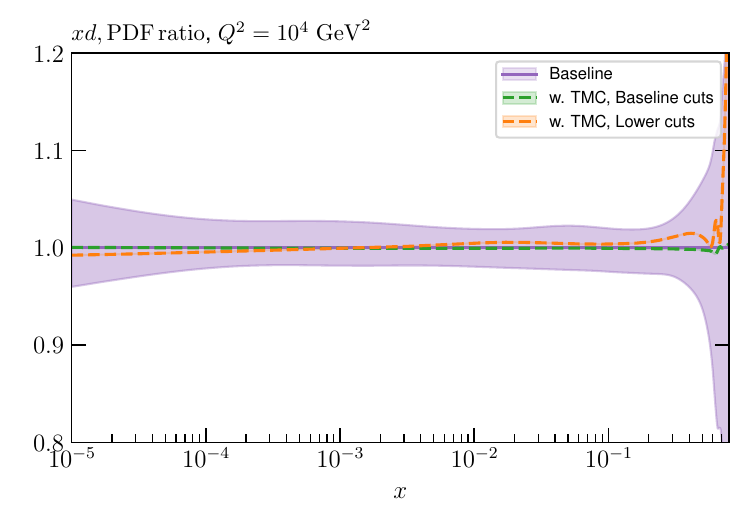}
\includegraphics[scale=0.5]{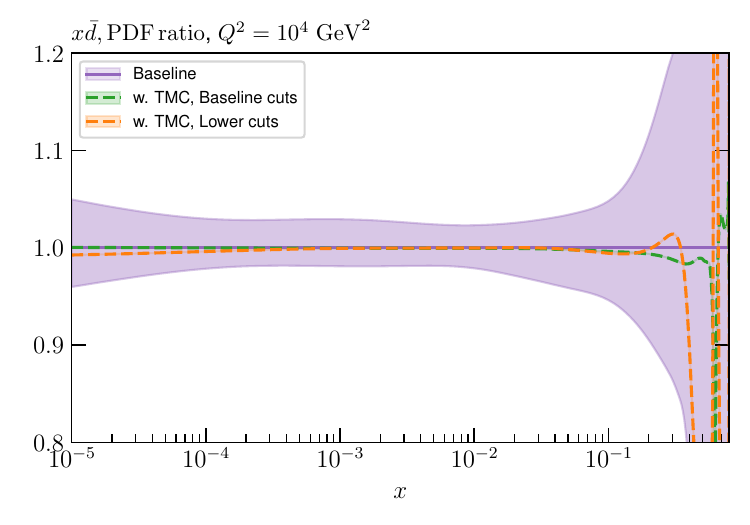}
\includegraphics[scale=0.5]{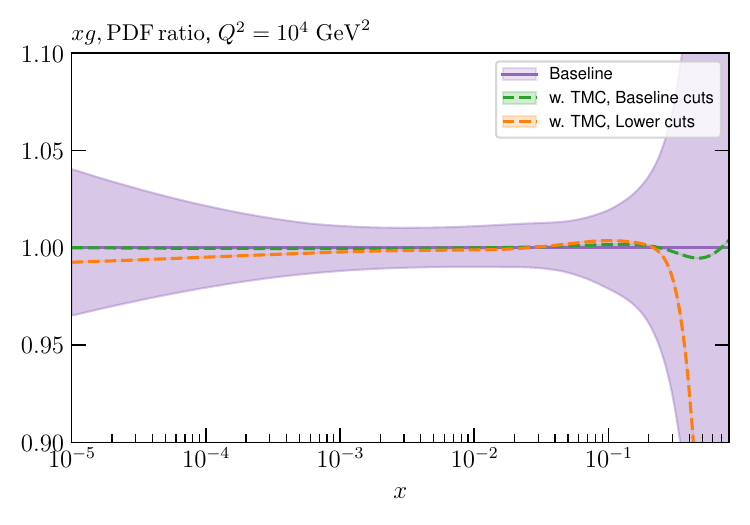}
\includegraphics[scale=0.5]{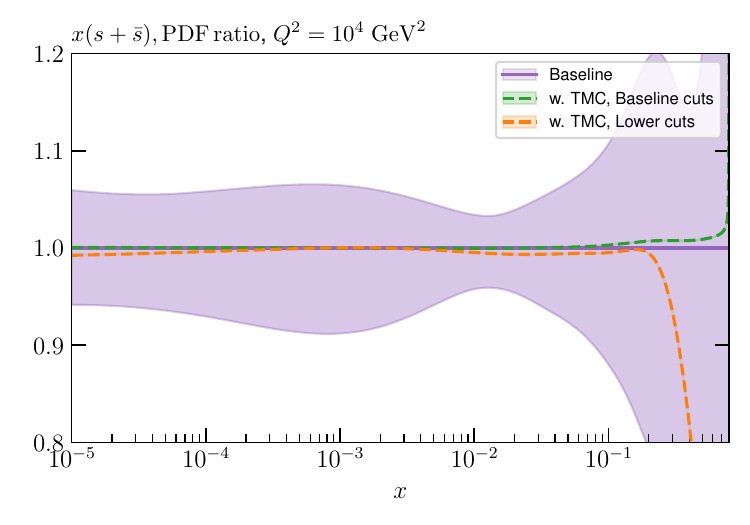}
\caption{\sf A selection of PDFs at $Q^2=10^4$ ${\rm GeV}^2$ that result from aN${}^3$LO MSHT fits with the new treatment of the fixed target datasets, and with target mass corrections included or excluded, and with a lower (default) $W^2$ cut of 5 (15) ${\rm GeV}^2$, as indicated.}
\label{fig:TMCcomp}
\end{center}
\end{figure}

We next consider the impact of these corrections on the PDFs, which is shown in Fig.~\ref{fig:TMCcomp}. The result with TMCs included on top of the baseline fit is shown for comparison, but as discussed above the sensitivity of this fit to these corrections is very minor, as is confirmed by the very small PDF impact. The impact of lowering the $W^2$ cut (and including TMCs) is larger but is in fact negligible for most PDFs and $x$ ranges. The dominant difference is, as expected, at higher $x$, notably in the $u,d$ quarks, which both show an enhancement. 

\begin{table}[t]
\begin{center}
  \scriptsize
  \centering
   \renewcommand{\arraystretch}{1.4}
\begin{tabular}{Xrcccc}\hline 
&Baseline cuts &  Baseline cuts &  Baseline cuts  & Lower $W^2$  \\
&(no HT, no TMC) & (w. HT, w. TMC) &(w. HT lower, w. TMC) &(w. HT, w. TMC)
\\ \hline
BCDMS $p$& 360.5 (1.10) & 331.6 (1.01) &367.0 (1.12)& 417.9 (1.19) \\
BCDMS $d$& 251.6 (1.02) & 234.0 (0.95) & 255.7 (1.04) & 298.4 (1.17) \\
NMC $p$ &383.9 (1.57) & 368.0 (1.51) &374.2 (1.53)& 405.7 (1.57) \\
NMC $d$&326.3 (1.34) & 295.9 (1.21)& 305.2 (1.25) & 335.0 (1.30)\\
SLAC $p$&31.3 (0.85)& 37.6 (1.02)&33.9 (0.92) & 101.3 (0.84) \\
SLAC $d$& 22.2 (0.58)& 27.4 (0.72) &19.1 (0.50) & 51.3 (0.44)\\
E665 $p$&  75.0 (1.41) & 78.1 (1.47)&77.7 (1.47) & 80.7 (1.52)\\
E665 $d$& 72.0 (1.36) & 77.8 (1.47) &75.6 (1.43) & 75.6 (1.43)\\
NuTeV $F_2$ &  35.6 (0.67)& 30.7 (0.58) &31.0 (0.58) & 35.2 (0.64)\\
NuTeV $F_3$ & 33.3 (0.79) & 31.4 (0.75) &30.6 (0.73) & 30.7 (0.73)\\
NMC $n/p$& 144.4 (0.98) & 142.3 (0.96)&141.5 (0.96) & 169.9 (0.97)\\
 \hline
{\bf Fixed Target}&{\bf 2201.1 (1.11)}  &{\bf 2116.0 (1.06)} & {\bf 2172.9 (1.09)} & {\bf 2464.3  (1.10)} 
\\ \hline
{\bf HERA}& {\bf 1626.9 (1.29)}& {\bf 1615.6 (1.28)} &{\bf 1620.7 (1.28)} &  {\bf 1618.8 (1.28)} \\
{\bf Hadron Collider}& {\bf 2386.8 (1.33)}& {\bf 2390.1 (1.33)}& {\bf 2381.9 (1.33)}&  {\bf 2385.7 (1.33)}
\\ \hline \hline 
{\bf Global }  &\bf{6214.7 (1.23)} &\bf{6121.7 (1.21)} &\bf{6175.5 (1.23)} &  \bf{6469.8 (1.22)}\\
\hline
\end{tabular}
\end{center}
\caption{\sf $\chi^2$ values for MSHT fits, with new treatment of the fixed target datasets, with higher twist corrections and  target mass corrections included, and with a lower (default) $W^2$ cut of 5 (15) ${\rm GeV}^2$, as indicated. In the baseline case, higher twist corrections are not included. The (HT lower) case corresponds to fixing the HT corrections at the best fit when the lower cut is applied. The absolute value is given, along with the $\chi^2$  per point in brackets, for the individual fixed target datasets, as well as for the results for the global dataset, and subsets of it. The lower (higher) $W^2$ cut corresponds to 2239 (1983) fixed target data points.}
\label{tab:chi2_HTcomp}
\end{table}

We next consider the impact of HT corrections. The corresponding fit qualities when these are allowed to be free are shown in Table~\ref{tab:chi2_HTcomp}. The impact of the inclusion of HT corrections for the lower cut can be seen by comparing the last column of this Table with that of Table~\ref{tab:chi2_TMCcomp}, and we can see this leads to an improvement of about 60 points in $\chi^2$, or $0.01$ per point globally, with this largely resulting from an improvement directly in the fit quality to the fixed--target data. This confirms the statement above that there is still a preference for a non--zero HT corrections in this case. 

However, we can also see a clear improvement in the fit quality for the baseline fit, even though in this case the higher $W^2$ cut is designed to reduce sensitivity to HT corrections. In particular, the fit quality improves by about 100 points in $\chi^2$, or roughly 0.02 per point globally. This is again largely driven again by a direct improvement in the fit quality to the fixed--target data, with the BCDMS and NMC datasets improving most in general. We will examine and compare the resulting HT corrections further below, but for now can make some initial comments regarding the fit quality. First, if we instead fix the HT corrections to the best fit values that come from the fit with the lower $W^2$ cut, and then refit the PDFs with the baseline cut, we can see from the Table that there is still some improvement in the fit quality, by about 40 point in $\chi^2$, but that this is much milder.  Moreover, if we attempt the converse, namely fixing the HT corrections from the fit with the baseline cut and performing a PDF fit with the lower $W^2$ cut, we find that the fit quality is terrible, with a global $\chi^2$ per point of $\sim 1.55$. This therefore indicates that the form of the preferred HT corrections is rather different in the two (lower and higher $W^2$ cut) cases; we will confirm this below.

\begin{table}
\begin{center}
  \scriptsize
  \centering
   \renewcommand{\arraystretch}{1.4}
\resizebox{\textwidth}{!}{\begin{tabular}{Xrccccc}\hline 
&Baseline  & Baseline cuts &  Baseline cuts  & Lower $W^2$& Lower $W^2$ \\
&(no HT, no TMC)&(w. HT, w. TMC)&(w. HT lower, w. TMC)&(no HT, w. TMC)&(w. HT, w. TMC)
\\ \hline
{\bf Fixed Target}& {\bf 2204.5 (1.11)} &{\bf 2125.0 (1.07)} &{\bf 2166.6 (1.09)}& {\bf 2541.3 (1.13)} & {\bf 2453.8 (1.09)} 
\\ \hline
{\bf HERA}& {\bf 1622.4 (1.28)}&{\bf 1602.9 (1.27)}&{\bf 1610.2 (1.27)}& {\bf 1627.3 (1.29)} & {\bf 1612.7 (1.28)} \\
{\bf Hadron Collider}& {\bf 2510.8 (1.40)} &{\bf 2492.8 (1.39)}& {\bf 2492.7 (1.39)}& {\bf 2519.2 (1.41)}& {\bf 2487.2 (1.39)}
\\ \hline \hline 
{\bf Global } &\bf{6337.7 (1.26)}  & \bf{6220.7 (1.23)}& \bf{6269.4 (1.24)}& \bf{6687.8 (1.26)}& \bf{6553.6 (1.24)}\\
\hline
\end{tabular}}
\end{center}
\caption{\sf As in Table~\ref{tab:chi2_HTcomp} but at NNLO, and only showing the results for the global fit quality, and the subsets of them. Also shown is the result for the lower $W^2$ cut with TMCs alone included.}
\label{tab:chi2_HTcomp_NNLO}
\end{table}

The corresponding results at NNLO are shown in Table~\ref{tab:chi2_HTcomp_NNLO}. The broad trends are similar to the aN${}^3$LO case, but with some differences. In particular, we can see that improvement from including HT corrections is somewhat larger. Moreover, for the baseline cuts the result of taking the HT corrections fixed to the best fit values from the lower cut fit is to lead to a rather larger improvement with respect to the baseline case. This all indicates that the preferred size of the HT corrections may be rather larger at NNLO.

\begin{figure}
\begin{center}
\includegraphics[scale=0.5]{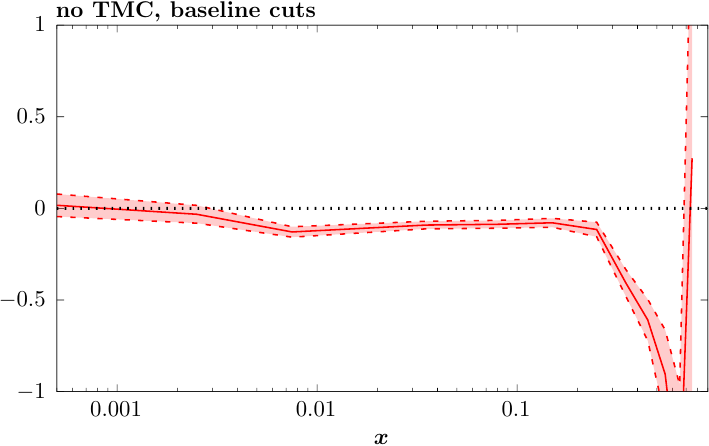}
\includegraphics[scale=0.5]{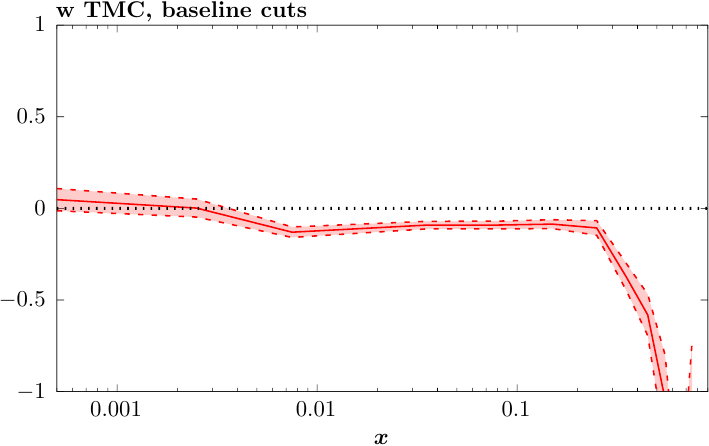}
\includegraphics[scale=0.5]{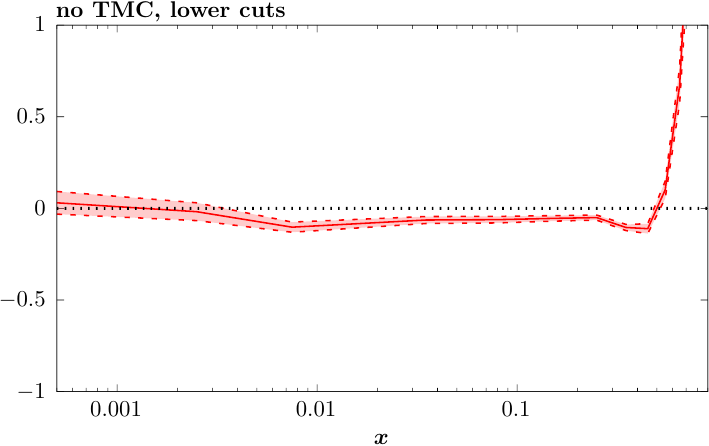}
\includegraphics[scale=0.5]{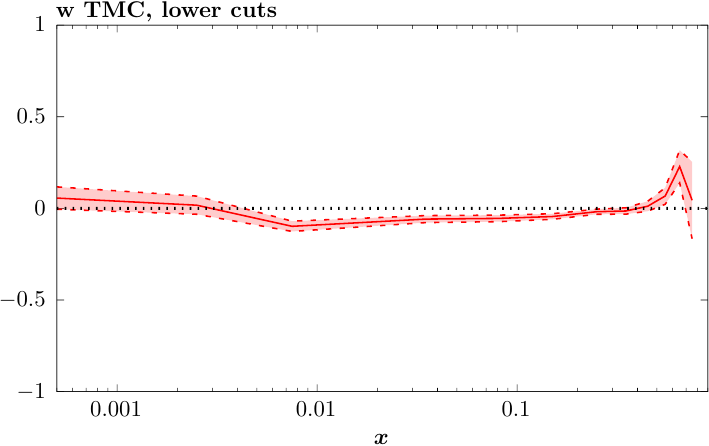}
\caption{\sf The size of the higher twist coefficients $a_i$ for fits with (left) and without (right) target mass corrections, and with (top) the baseline and (bottom) the lower $W^2$ cut for the aN$^3$LO fit. The red band indicates the diagonal $\Delta \chi^2=1$ uncertainty on these from the fit, for demonstration purposes.}
\label{fig:HTs}
\end{center}
\end{figure}

\begin{figure}
\begin{center}
\includegraphics[scale=0.5]{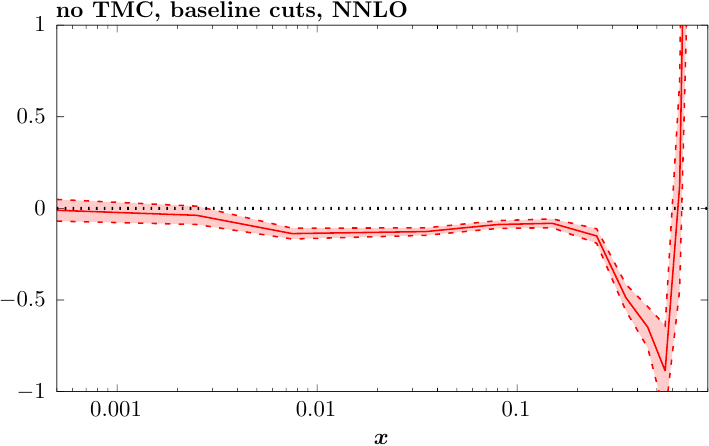}
\includegraphics[scale=0.5]{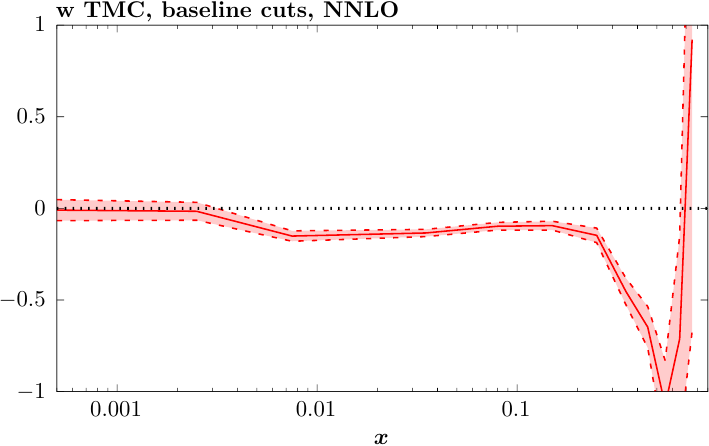}
\includegraphics[scale=0.5]{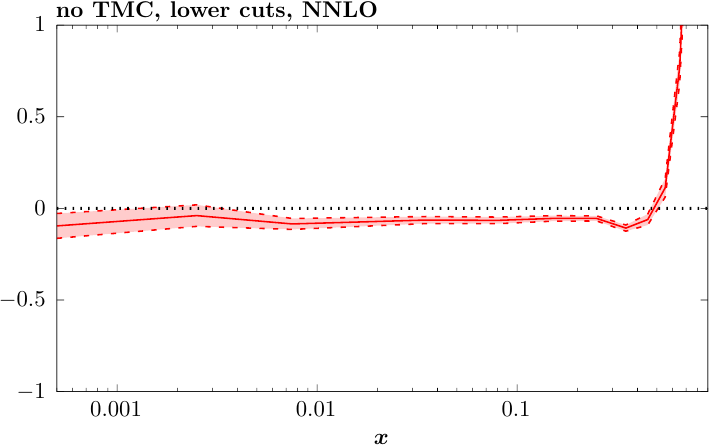}
\includegraphics[scale=0.5]{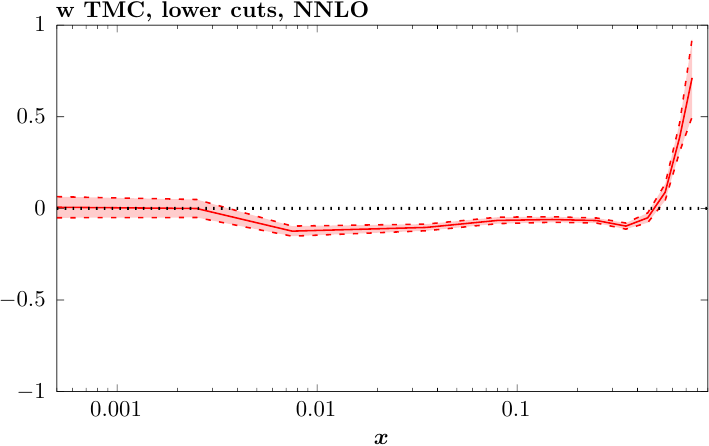}
\caption{\sf As in Fig.~\ref{fig:HTs} but at NNLO.}
\label{fig:HTs_NNLO}
\end{center}
\end{figure}

Turning now to the size of the HT corrections, these are shown in Figs.~\ref{fig:HTs} and~\ref{fig:HTs_NNLO} at aN${}^3$LO and NNLO, respectively. Broadly we can see a well established trend seen in earlier studies, namely that the larger relative HT corrections are rather concentrated at high $x$. It should be noted however that the structure function, $F_2$, that these HT multiply are of course steeply falling functions with $x$ towards the end point region, and hence in absolute terms the corrections are more evenly spread in $x$. Starting with the left plots, these exclude TMCs, for the sake of comparison. In other words, these examine the extent to which the freely parameterised HT corrections can absorb TMCs, if these are not explicitly included. We can see that the free HT coefficients prefer rather steeply increasing corrections towards high $x$, which are precisely in line with the expectations from the known TMCs. Interestingly, even for the baseline case some tendency for this increase is seen at the highest $x$ values, indicating that while the actual impact on the fit quality from TMCs is small for the baseline cuts, there are clearly some kinematic bins in the fixed--target dataset that show some preference for TMCs.

Including the explicit calculation of TMCs, we are then left with the remaining HT contribution, which are shown in the right plots. We can see that in terms of the size of these, they are clearly smaller than the left hand case, consistent with the fact that the dominant improvement in the fit quality in the lower $W^2$ cut case comes from the inclusions of TMCs. Comparing the top and bottom right plots, we can see a clear difference in this $x$ behaviour between the baseline and lower $W^2$ cut cases. For the lower $W^2$ cut, the HT correction increases towards the highest $x$ values, albeit with a dip in the final bin. On the other hand, for the baseline cut case, the trend is reversed, with the correction being rather larger and negative at higher $x$. This difference in behaviour is therefore indeed consistent with the expectations based on the fit quality above, namely that the preferred HT corrections for the lower cut are not consistent with those from the baseline case; we will discuss this further below.

In the NNLO case shown in Fig.~\ref{fig:HTs_NNLO}, the trend in the left plots is broadly similar to the aN${}^3$LO case, namely again indicating the dominance of TMCs in the overall correction. Considering the right plots, we can see that the trend for the remaining HT corrections is  similar, but not identical. In particular, for the lower $W^2$ cut, we can see that the increase of the HT corrections with higher $x$ is more significant, i.e. the HT correction is larger (a similar effect was seen in the earlier study~\cite{Blumlein:2006be}). In the baseline cut case, while there is evidence of a clear negative dip towards higher $x$, we can see that at the highest $x$ values there is a trend or the correction to increase.

\begin{figure}
\begin{center}
\includegraphics[scale=0.5]{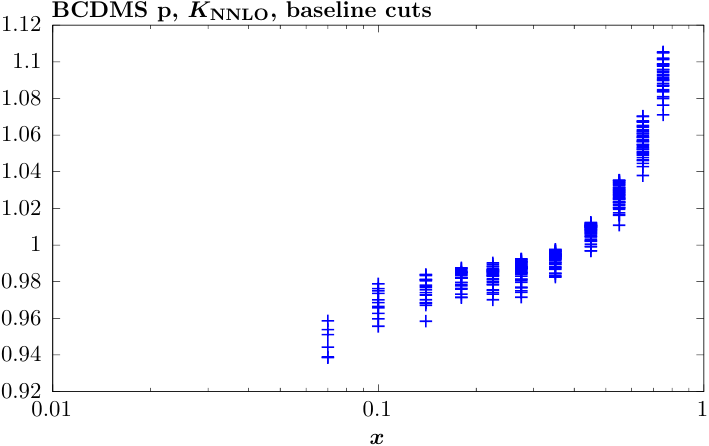}
\includegraphics[scale=0.5]{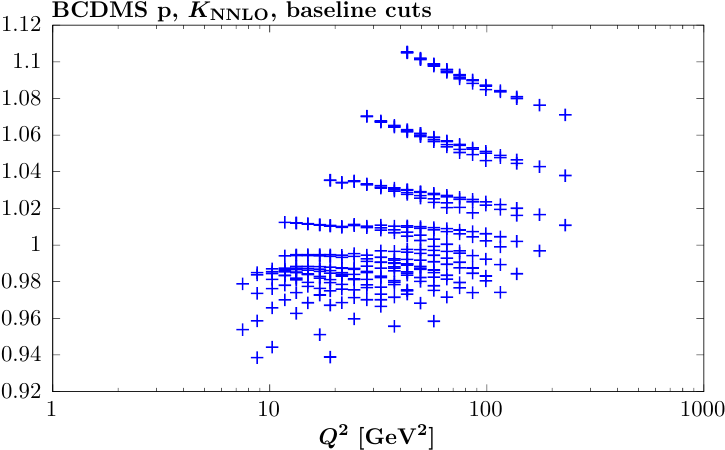}
\includegraphics[scale=0.5]{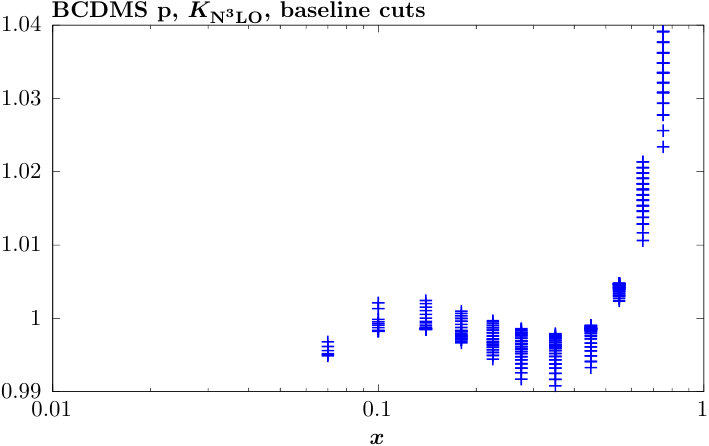}
\includegraphics[scale=0.5]{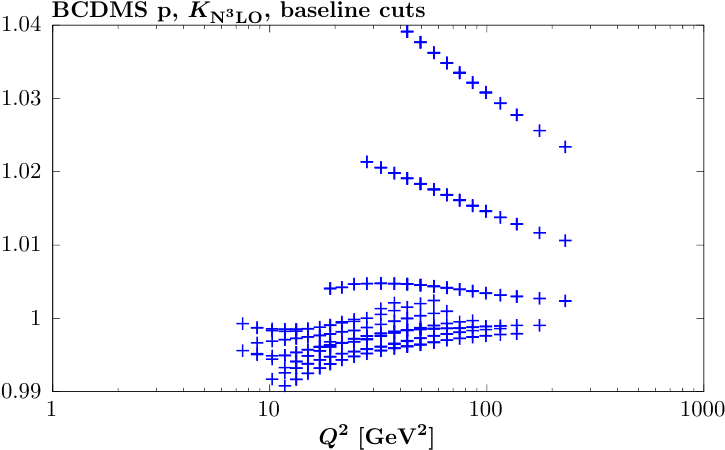}
\includegraphics[scale=0.5]{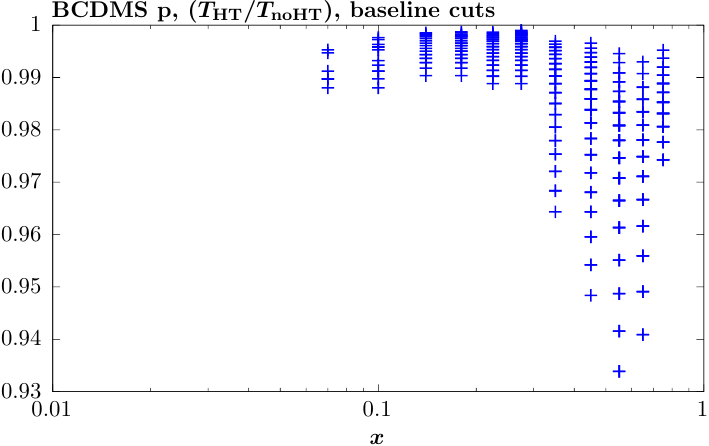}
\includegraphics[scale=0.5]{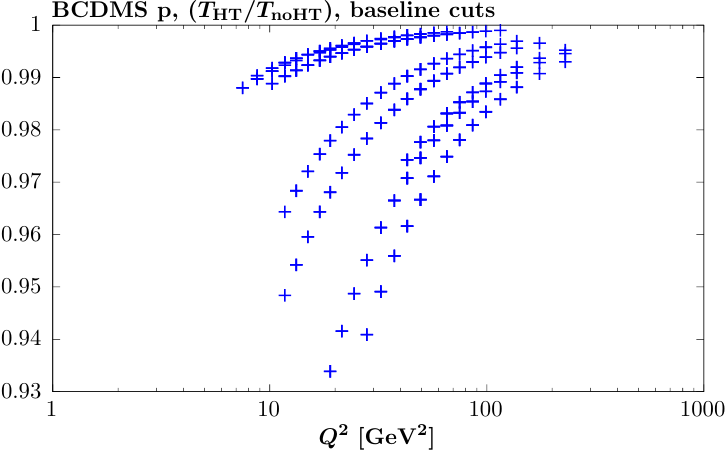}
\includegraphics[scale=0.5]{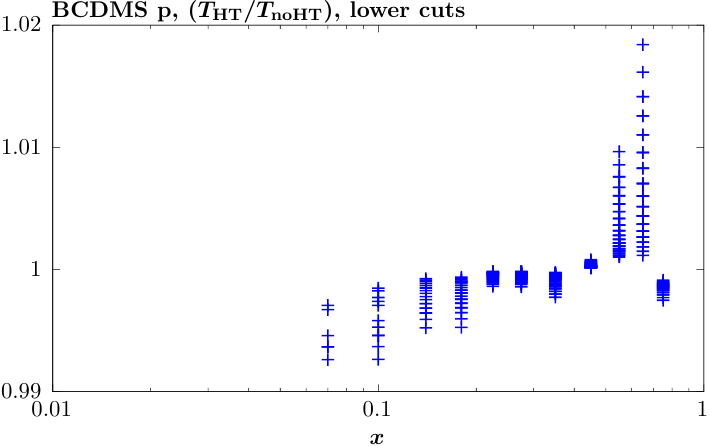}
\includegraphics[scale=0.5]{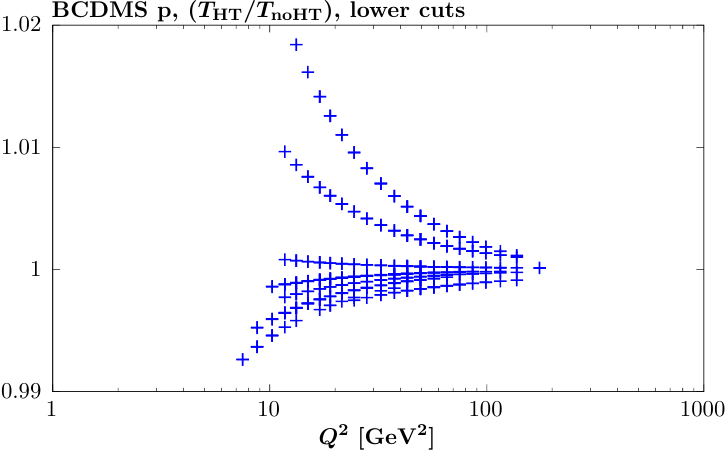}
\caption{\sf Different theoretical corrections for the BCDMS proton data vs. $x$ (left) and $Q^2$ (right). The upper (second) panel shows the NNLO (N${}^3$LO) QCD K--factor, while the third (final) panel shows the size of the HT correction for the baseline (lower $W^2$) cuts, in the N${}^3$LO fit.}
\label{fig:HTcompBCDMS}
\end{center}
\end{figure}

To understand these trends further, in Fig.~\ref{fig:HTcompBCDMS} we show both the NNLO and  N${}^3$LO QCD K--factors for the baseline PDF set for a relevant example of the BCDMS proton data, alongside the HT corrections that come from the aN$^3$LO fit with the baseline and lower $W^2$ cuts. From this, a number of comments can be made. First, comparing the third and bottom panels we can clearly see that the size and most notably trend of the HT corrections is very different, with the baseline case leading to a negative correction, consistently with Fig.~\ref{fig:HTs}. We can see in particular that even for relatively large $Q^2$ this correction is not negligible for the baseline cuts. For the lower $W^2$ cut, on the other hand, the correction is smaller, and for moderate to large $Q^2$ is generally permille level, as we may hope for from genuinely HT corrections. 

Looking at the upper panel, some explanation for this trend is suggested. In particular, we can see that the baseline HT corrections are very similar in size, and generally opposite in sign to, the N${}^3$LO QCD K--factors. It was discussed in \cite{Martin:2003sk} that at high $x$ the perturbative expansion of structure functions becomes unstable (as applied in the renormalon calculations of \cite{Dokshitzer:1995qm,Dasgupta:1996hh}), and at each $x$ value there is a region of $Q^2$ where a particular high order in the perturbative expansion is of the same size as the ambiguity in the sum of the perturbative expansion. Therefore, considering that remaining higher order corrections may be of the same order as the N${}^3$LO QCD K--factors, this indicates that the `HT' corrections that are fit in the baseline case are in fact not of dominantly higher twist origin, in the sense that they are genuine $1/Q^2$ corrections, but rather that the parameterisation of \eqref{eq:HT} is effectively absorbing the unknown remaining QCD corrections, and can do this rather effectively over the $Q^2$ region spanned by the fixed--target data. In particular, over a quite limited region of $Q^2$, corrections which behave like high powers of $\alpha_S(Q^2)$ can be effectively mimicked by terms behaving like $1/Q^2$. On the other hand, when the $W^2$ cut is lowered, genuine HT corrections become more significant, and these tend to dominate the fit to the HT parameterisation.  

\begin{figure}
\begin{center}
\includegraphics[scale=0.5]{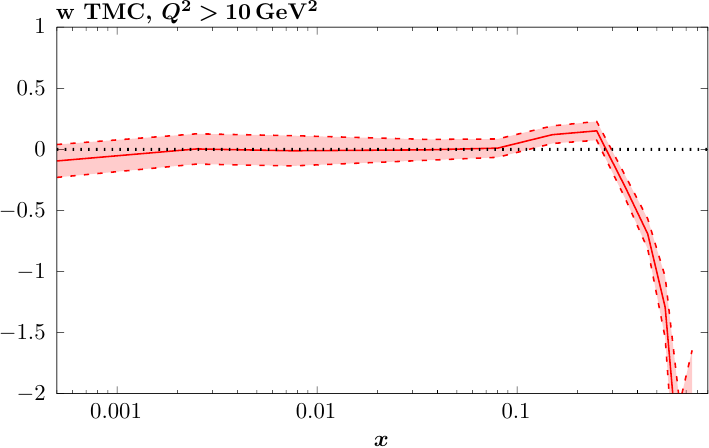}
\includegraphics[scale=0.5]{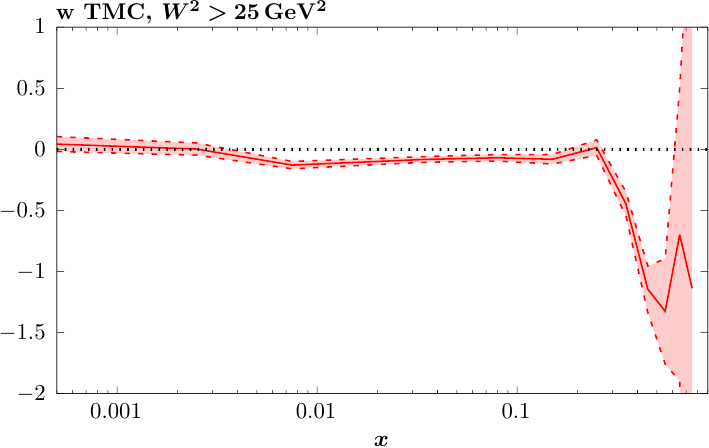}
\caption{\sf As in Fig.~\ref{fig:HTs} but for other cut selections.}
\label{fig:HTs_N3LO_othercuts}
\end{center}
\end{figure}

More precisely, one can expect both genuine HT and higher order QCD corrections to be to some extent absorbed into the HT parameterisation of \eqref{eq:HT}, with the precise amount depending on the cuts considered. Indeed, this is backed up by the previous discussion of the fit quality in Table~\ref{tab:chi2_HTcomp}, where we found that for the baseline cuts, taking HT corrections fit to the lower $W^2$ cut case, i.e. which may more reasonably be taken to be of purely HT origin, still gives a notable (but milder) improvement in the fit quality. With this in mind, in Fig.~\ref{fig:HTs_N3LO_othercuts} we show the HT corrections that come from fits with different cuts to the fixed--target (and now DIS HERA) data. Namely, we consider raising the $Q^2$ cut to 10 ${\rm GeV^2}$ or the $W^2$ cut to 25 ${\rm GeV^2}$. In this way the contribution from genuine HT corrections should be even further reduced. We can see that in both cases, and for the higher $Q^2$ cut in particular, the trend for a negative correction at high $x$ is further enhanced, supporting the point that one can readily absorb essentially purely higher order QCD corrections into the freedom of the HT parameterisation. We note that in terms of the fit quality, we again see a notable improvement upon the inclusion of `HT' corrections for these cuts.

\begin{figure}
\begin{center}
\includegraphics[scale=0.5]{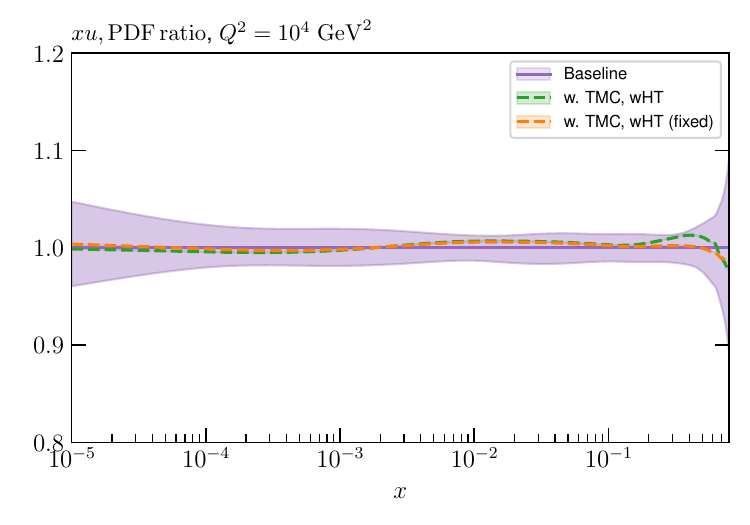}
\includegraphics[scale=0.5]{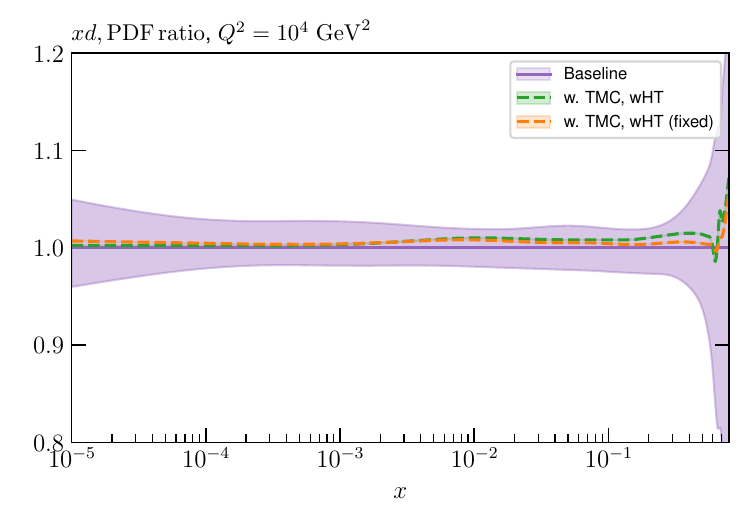}
\includegraphics[scale=0.5]{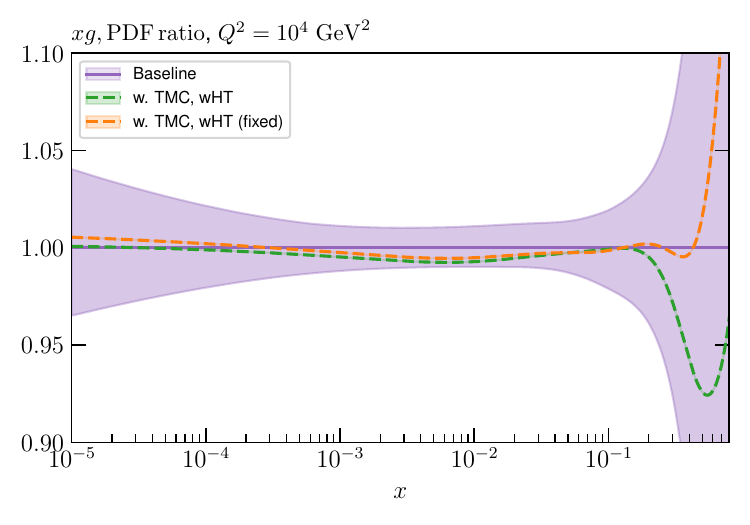}
\caption{\sf A selection of PDFs at $Q^2=10^4$ ${\rm GeV}^2$ that result from ${\rm aN}^3$LO MSHT fits, with higher twist corrections included or excluded, as indicated, for the baseline MSHT cuts. The HT (fixed) case corresponds to fixing these HT corrections to the result of the fit with the lower $W^2$ cuts. Results are shown for PDFs where the impact is largest.}
\label{fig:HTcompMSHTcuts_rat}
\end{center}
\end{figure}

\begin{figure}
\begin{center}
\includegraphics[scale=0.5]{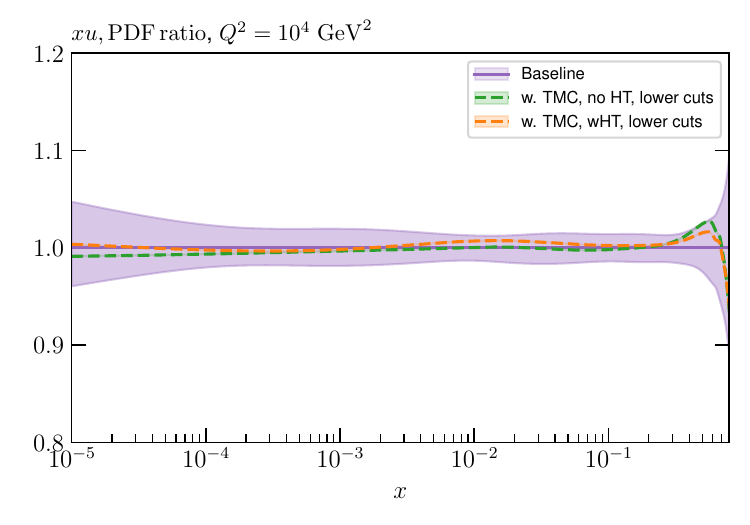}
\includegraphics[scale=0.5]{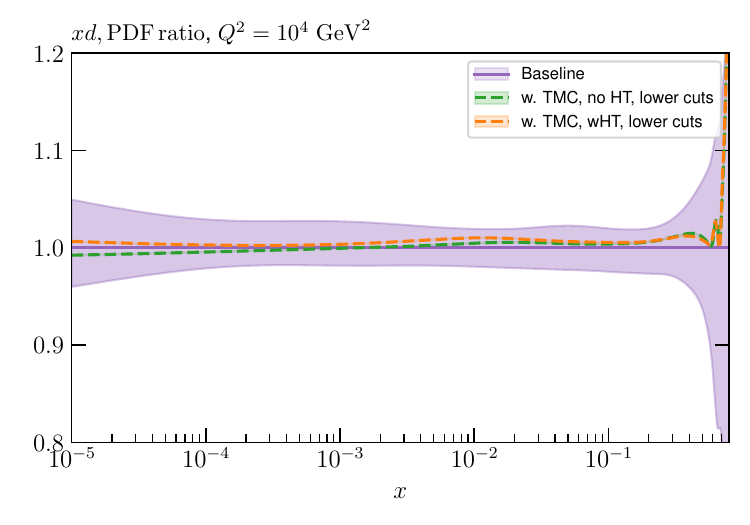}
\includegraphics[scale=0.5]{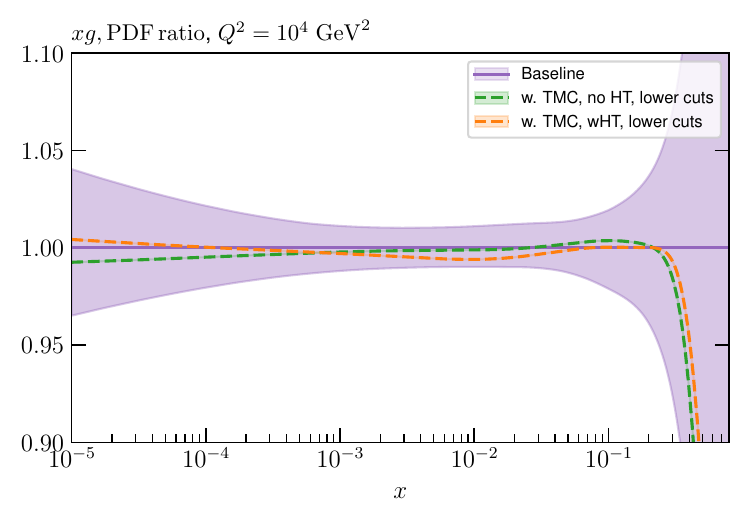}
\caption{\sf A selection of PDFs at $Q^2=10^4$ ${\rm GeV}^2$ that result from ${\rm aN}^3$LO MSHT fits, with higher twist corrections included or excluded, as indicated, for the lower $W^2$ cut. Results are shown for PDFs where the impact is largest.}
\label{fig:HTcompTMCcuts_rat}
\end{center}
\end{figure}

We next consider the impact on the PDFs, which is shown in Figs~\ref{fig:HTcompMSHTcuts_rat} and~\ref{fig:HTcompTMCcuts_rat} for the baseline and lower $W^2$ cut cases, at ${\rm aN}^3$LO. For the baseline case the impact of HTs, with these allowed to be fully free in the fit and fixed to best fit value from the lower $W^2$ cut is shown. In both cases the impact is generally very small, although there a few PDFs, namely the gluon and down quark at intermediate  to high $x$ and the up quark at high $x$ where the central PDFs from the HT fits lie towards the edge of the default uncertainty band, as calculated using the standard MSHT dynamic tolerance procedure. For the lower cut case, the trend is very similar indeed when HT corrections are included. We also show the result from a fit with only TMC corrections for comparison, although given additional HT corrections are clearly important for such a fit, this is not a realistic result. Nonetheless, this shows that the impact of TMCs alone on the resulting PDFs is similar.

\begin{figure}
\begin{center}
\includegraphics[scale=0.5]{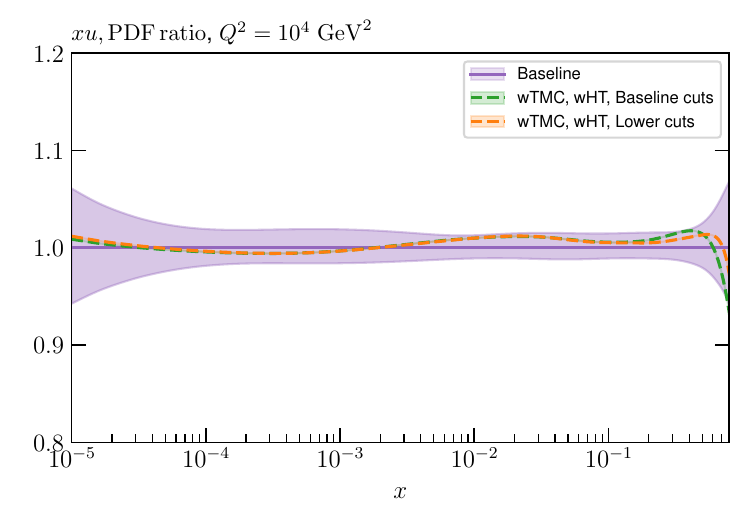}
\includegraphics[scale=0.5]{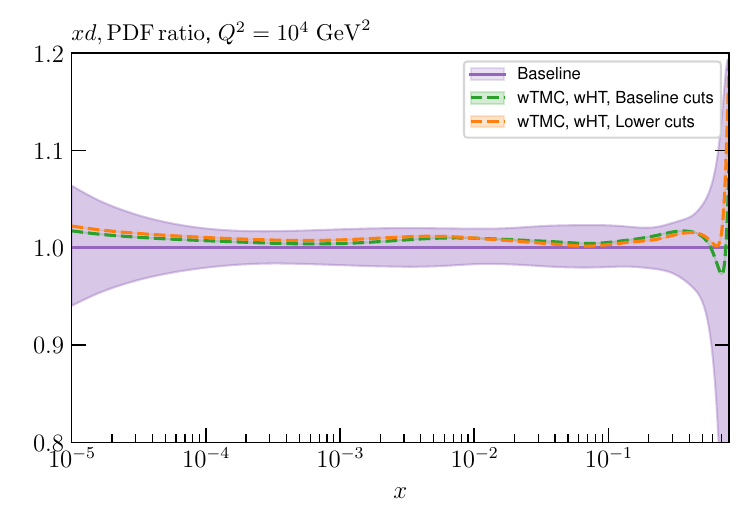}
\includegraphics[scale=0.5]{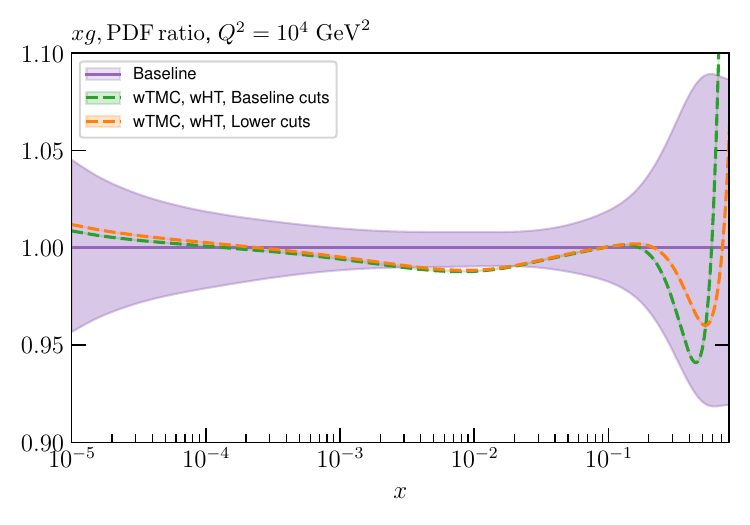}
\caption{\sf As in Fig.~\ref{fig:HTcompTMCcuts_rat} but at NNLO.}
\label{fig:HTcompTMCcuts_NNLO}
\end{center}
\end{figure}

\begin{figure}
\begin{center}
\includegraphics[scale=0.5]{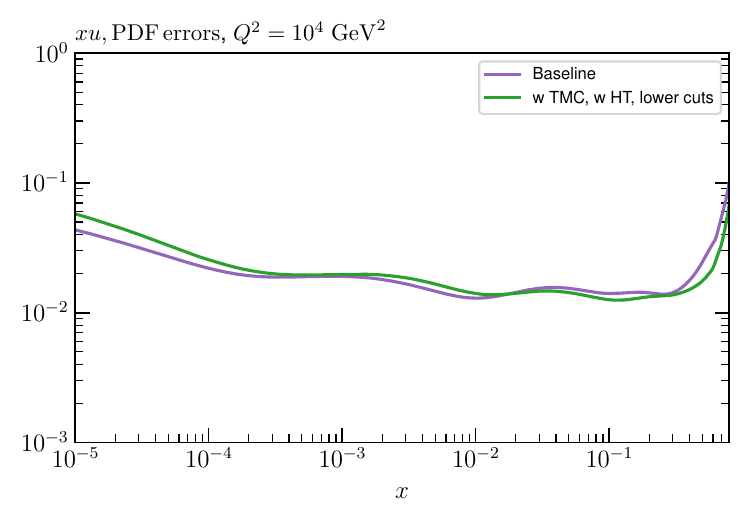}
\includegraphics[scale=0.5]{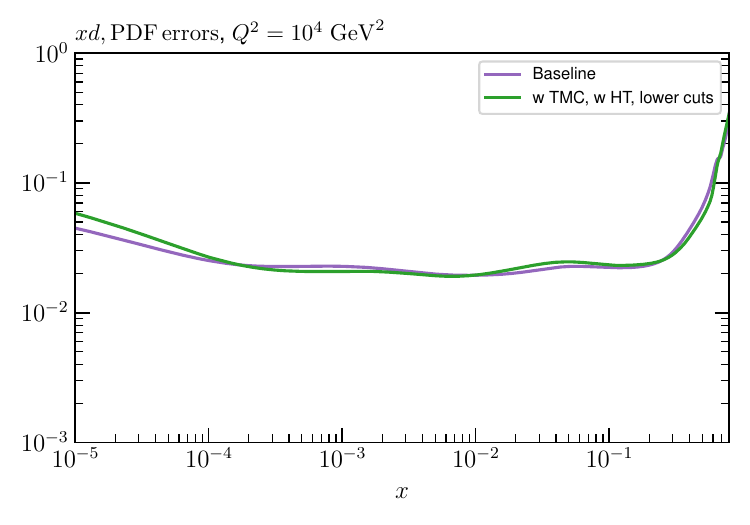}
\includegraphics[scale=0.5]{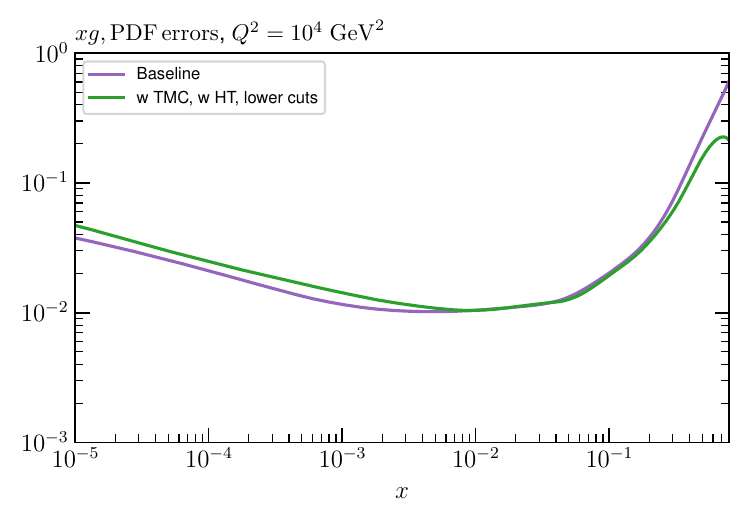}
\caption{\sf PDF uncertainties for a selection of PDFs at $Q^2=10^4$ ${\rm GeV}^2$ that result from ${\rm aN}^3$LO MSHT fits with the new treatment of the fixed target datasets, with the baseline cuts, excluding higher twist and TMC corrections and a lower $W^2$ cut, including higher twist and TMC corrections. Results are shown for the PDF selection considered in Fig.~\ref{fig:HTcompTMCcuts_rat}}
\label{fig:cutcomp_err}
\end{center}
\end{figure}

\begin{figure}
\begin{center}
\includegraphics[scale=0.5]{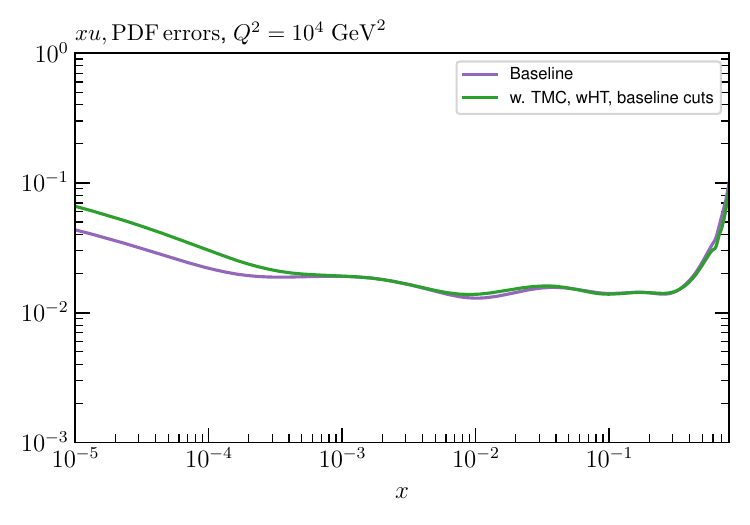}
\includegraphics[scale=0.5]{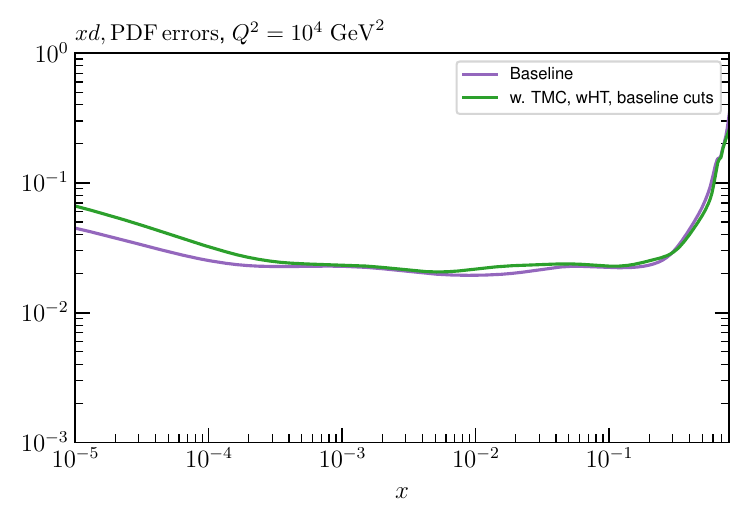}
\includegraphics[scale=0.5]{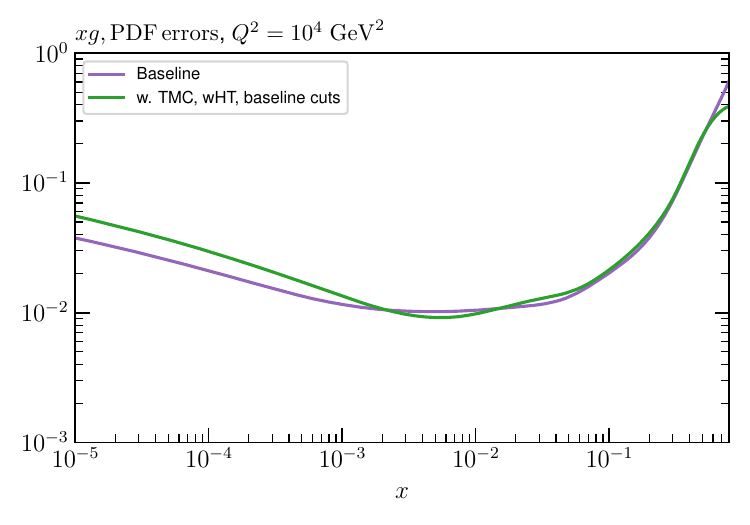}
\caption{\sf PDF uncertainties for a selection of PDFs at $Q^2=10^4$ ${\rm GeV}^2$ that result from  ${\rm aN}^3$LO  MSHT fits with the new treatment of the fixed target datasets, with the baseline cuts (excluding higher twist and target mass corrections), and with target mass and/or higher twist corrections included. Results are shown for the PDF selection considered in Fig.~\ref{fig:HTcompTMCcuts_rat}}
\label{fig:HTcompMSHTcuts_err}
\end{center}
\end{figure}

The impact of reducing the $W^2$ cut on the PDF uncertainties is shown in Fig.~\ref{fig:cutcomp_err}. This is  rather mild; if higher twist corrections are excluded (not shown) then the result is rather similar. The impact of including HT corrections in the baseline cut case is shown in Fig.~\ref{fig:HTcompMSHTcuts_err}, and is seen to give some increase in the PDF uncertainty at low $x$.

Finally, we consider the NNLO case, with the equivalent PDF impacts to the ${\rm aN}^3$LO case in Fig.~\ref{fig:HTcompTMCcuts_rat} is shown in Fig.~\ref{fig:HTcompTMCcuts_NNLO}. We can see that the broad trends are very similar, but interestingly the size of the impact of including HT corrections is rather larger in this case, which is in line with the larger HT corrections  seen in Fig.~\ref{fig:HTs_NNLO}. 

\begin{figure}
\begin{center}
\includegraphics[scale=0.5]{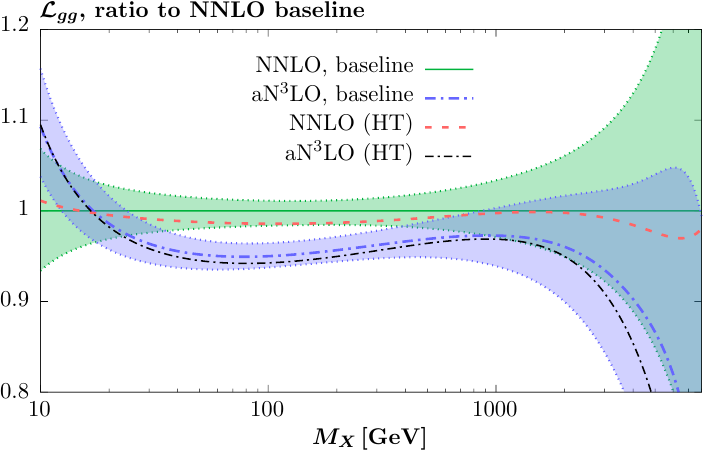}
\caption{\sf Ratio of the gluon--gluon luminosities to the baseline NNLO case, for MSHT fits at both NNLO and  ${\rm aN}^3$LO. The `HT' case corresponds to fits including both TMCS and HT corrections, and with the lower $W^2$ cut imposed.}
\label{fig:gglumi}
\end{center}
\end{figure}

One particular effect of interest is the decrease in the gluon PDF in the $x$ region relevant for Higgs boson production in gluon fusion in both the NNLO and aN$^3$LO fits. In Fig.~\ref{fig:gglumi} we show the corresponding gluon--gluon luminosities and this reduction is again clear in the Higgs mass region. It is in particular of note that the level of reduction in the NNLO case is somewhat larger, which will therefore act to stabilise, albeit rather mildly, the difference between the NNLO and ${\rm aN}^3$LO cases.

\begin{figure}
\begin{center}
\includegraphics[scale=0.5]{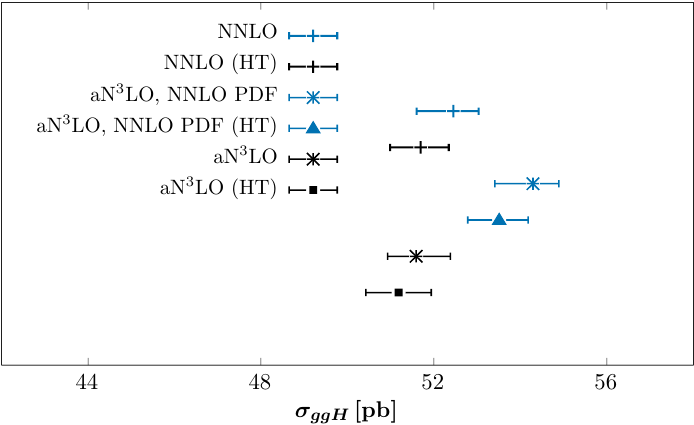}
\includegraphics[scale=0.5]{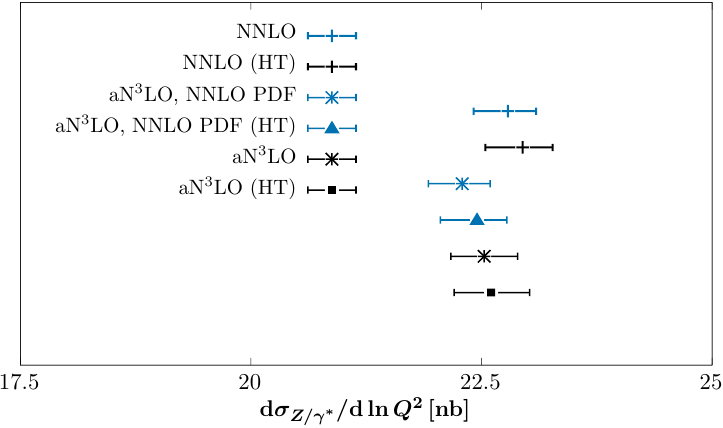}
\includegraphics[scale=0.5]{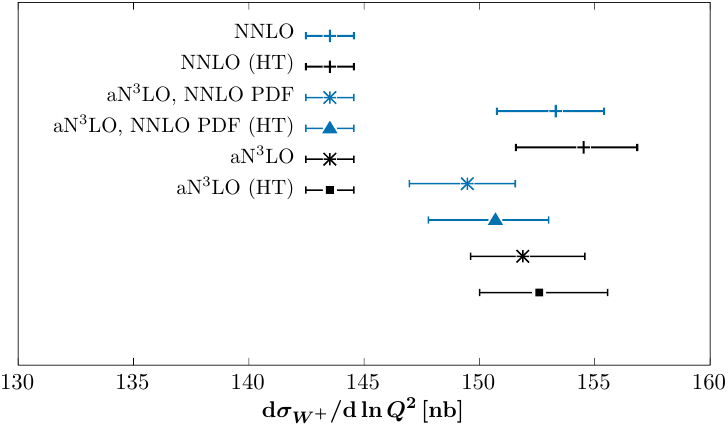}
\includegraphics[scale=0.5]{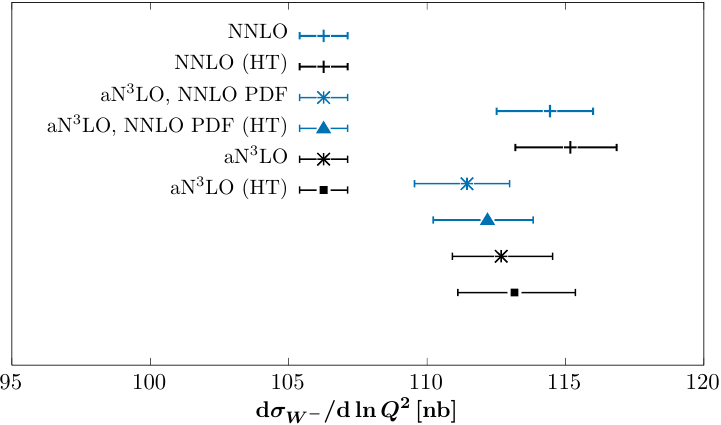}
\caption{\sf Cross section predictions for $ggH$, $Z$ and $W^\pm$ production at the $\sqrt{s}=14$ TeV LHC, for different MSHT fits, as indicated by the legends. Preditions are calculated using the \texttt{n3loxs}~\cite{Baglio:2022wzu} and as described in~\cite{Cridge:2023ryv}. PDF uncertainties alone are shown.}
\label{fig:xs}
\end{center}
\end{figure}

To investigate this, and the broader phenomenological impact of including HT corrections, in Fig.~\ref{fig:xs} we show cross section predictions for $ggH$, $Z$ and $W^\pm$ production at the $\sqrt{s}=14$ TeV LHC, for different MSHT fits, as indicated by the legends. Predictions are calculated using the \texttt{n3loxs}~\cite{Baglio:2022wzu} and as described in~\cite{Cridge:2023ryv}. In the Higgs case, we can see as expected a reduction in the predicted cross section is observed at both NNLO and ${\rm aN}^3$LO, but indeed with some  reduction in the difference between results with the two different PDFs applied observed. This effect is nonetheless clearly quite mild, and is certainly well within PDF uncertainties, and indeed other uncertainties due e.g. to scale variation, which for simplicity we do not show. The changes in the $Z$ and $W^\pm$ cases are generally somewhat smaller, although still visible, with the inclusion of HT corrections tending to increase the predicted cross sections.

\subsection{Impact on Strong Coupling}\label{sec:alphas}

In this section we consider the impact of the above corrections on the preferred value of the strong coupling. The extent to which HT corrections impact on this has in particular been a topic of debate~\cite{Blumlein:2006be,Alekhin:2012ig,Thorne:2014toa}, with some studies observing a significant downwards shift in the preferred value of the strong coupling when HT corrections are included, although the interpretation of this is tied up with whether a fixed or variable heavy flavour scheme used.

\begin{figure}
\begin{center}
\includegraphics[scale=0.8]{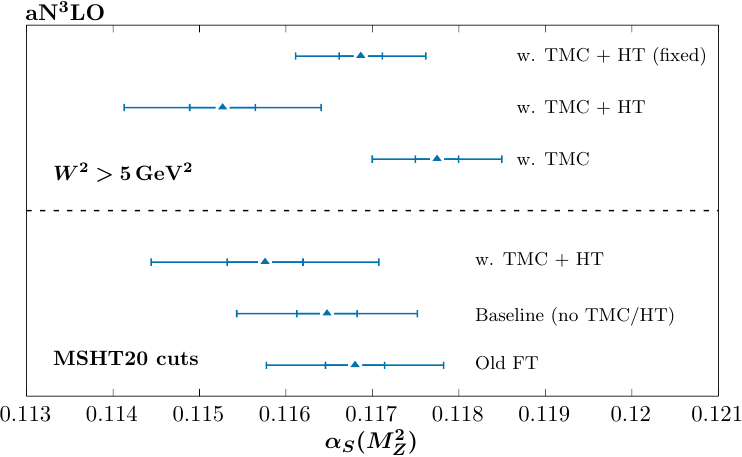}
\caption{\sf Best fit value of the strong coupling $\alpha_S(M_Z^2)$ for a range of aN${}^3$LO fits. The inner (outer) limits correspond to $\Delta \chi^2=1$ (9) that result from a quadratic fit about the minimum to the corresponding global profile, and are intended to be indicative of the relative uncertainty on $\alpha_S(M_Z^2)$ in different fits, rather than indicating the actual size of the uncertainty that would result from the dynamic tolerance procedure. }
\label{fig:as_comps_n3lo}
\end{center}
\end{figure}

In Fig.~\ref{fig:as_comps_n3lo} we show the results for the best fit $\alpha_S(M_Z^2)$ values and an estimate of the extent to which they are constrained for a range of aN${}^3$LO fits. In particular, a scan over fixed values of $\alpha_S(M_Z^2)$, spaced by 0.001, is performed in each case and the three points closest to the minimum are used for a quadratic fit to the $\Delta \chi^2$ profile. In the figure the inner (outer) limits correspond to the  $\Delta \chi^2=1$ (9) cases, and are shown for guidance. In particular, the true uncertainty in any of these fits would be calculated using the dynamic tolerance criterion (see~\cite{Cridge:2024exf} for discussion in the context of a recent fit to the strong coupling), which generally corresponds to an uncertainty that is larger than the outer  $\Delta \chi^2=9$ limit, though it is of that order. 

\begin{figure}
\begin{center}
\includegraphics[scale=0.6]{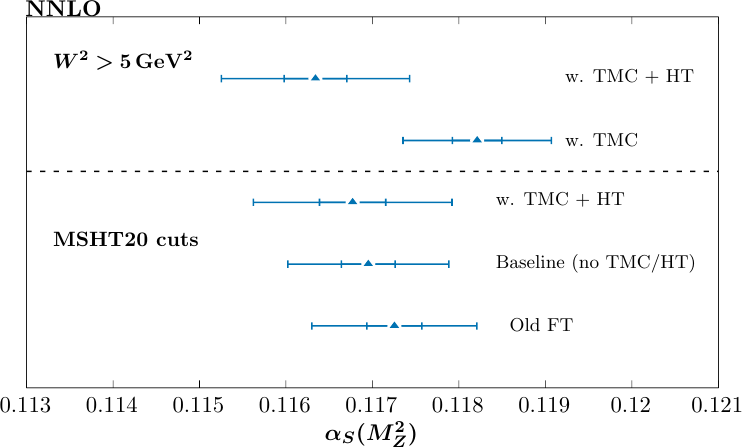}
\caption{\sf As in Fig.~\ref{fig:as_comps_n3lo} but for NNLO fits.}
\label{fig:as_comps_nnlo}
\end{center}
\end{figure}

Starting with the baseline (MSHT20) cuts, we can see the impact of the updated fixed--target data treatment on the strong coupling by comparing the bottom and second from bottom cases. We in particular find that this update leads a moderate reduction in the best fit value, consistent with the fact that this tends to increase the constraining power of the fixed--target data, which itself generally prefers lower values of the strong coupling. We note that the baseline (`Old FT') value is  slightly lower than the best fit found in~\cite{Cridge:2024exf}, despite the fact that the datasets in the fits are very similar; the central value has in particular dropped from 0.1170 to 0.1168. The reason for this in part related to the updated aN${}^3$LO evolution but also due to a corrected treatment of the $\alpha_S$ dependence of the aN${}^3$LO K-factors for LHC jet, Drell Yan and $Z$ $p_\perp$ data. In fact,  the inclusion of the Seaquest data tends to increase the preferred value of the strong coupling at this order, which counteracts the effects above.  The overall shift is therefore very small, and is certainly well within the quoted uncertainty in~\cite{Cridge:2024exf} of $\pm 0.0016$, though as we can see is further enhanced by the update of the fixed--target data. 

\begin{figure}
\begin{center}
\includegraphics[scale=0.5]{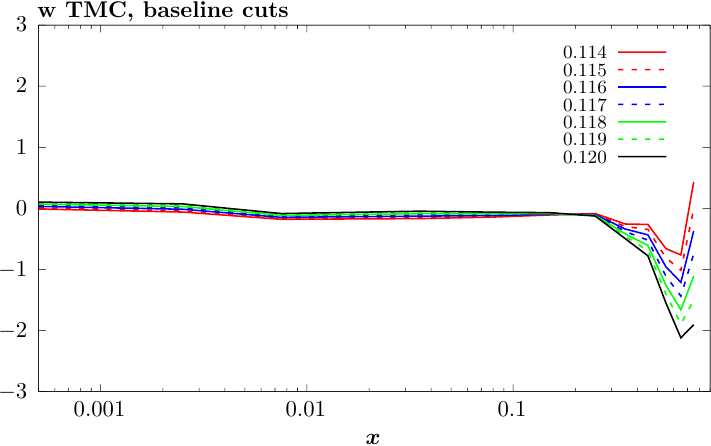}
\includegraphics[scale=0.5]{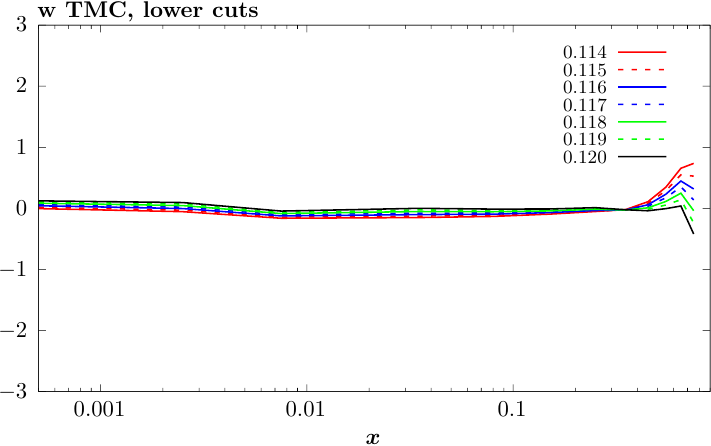}
\caption{\sf The higher twist corrections corresponding to Figs.~\ref{fig:as_comps_nnlo} and~\ref{fig:as_comps_n3lo}, as indicated above the figures.}
\label{fig:as_N3LO_HTs}
\end{center}
\end{figure}

The NNLO results are shown in Fig.~\ref{fig:as_comps_nnlo} and we can see a similar level of shift downwards occurs upon updating the fixed--target dataset treatment between the two orders, with the aN${}^3$LO case preferring a lower value to NNLO. We note again the difference between the two orders is somewhat different to that observed in~\cite{Cridge:2024exf}, and can be explained by the effects discussed above. The direction and size of this difference between NNLO and aN${}^3$LO is consistent with that recently observed in~\cite{Ball:2025xgq}, although the baseline central values at both orders are at slightly higher values then found here.

Turning now to the impact of HT (and TMC) corrections, at  aN${}^3$LO we can see that for the baseline cuts these lead to something of a reduction in the central value, by about $\sim 0.007$. The corresponding HT corrections for different $\alpha_S(M_Z^2)$ values is shown in Fig.~\ref{fig:as_N3LO_HTs} (left) and we can see that no particularly unusual behaviour is preferred for the lower values, which in fact prefer a somewhat smaller HT (or more properly HT + missing higher order QCD) correction at higher $x$; this is most likely due to the fact that for smaller coupling values the  higher order QCD corrections are reduced and hence a smaller HT correction is preferred to match the data. We have also investigated the effect of fixing the  aN${}^3$LO K-factors and/or HT corrections  to the best fit values at $\alpha_S(M_Z^2)$ of 0.118. This leads to a mildly higher central value, but the effect is rather marginal, and we therefore do not include these in the figure. In other words, the additional freedom from the missing higher order corrections to hadronic processes and HT corrections does not dramatically change the picture. The latter point is particularly relevant, as it indicates that the freedom to refit the HT corrections with $\alpha_S(M_Z^2)$ is not playing a major role in reducing the best fit value. Overall however, the size of the shift is well within the $T=3$ uncertainty, which is of the order of the result of the full dynamic tolerance procedure, but is at the edge of the textbook $T=1$ uncertainty. 

We next consider the case with the lower $W^2$ cut. The result with TMCs alone included is shown, although only for indication as we expect there to be additional, potentially important HT corrections in such a fit, which should be accounted for to give a reliable result. In this case, we can see that the result with purely TMCs prefers a rather higher value of the coupling than the baseline fit, being $\sim 0.0012$ higher. This is of the order of the dynamic tolerance uncertainty, and well outside the $T=1$ case. On the other hand, adding in HT corrections leads to a rather larger reduction, by $\sim 0.0024$, with respect to the TMC alone case, and $\sim 0.005$ with respect to the baseline cuts case, including TMC + HT corrections. Interestingly now, however, the impact of fixing the HT corrections to the best fit values at $\alpha_S(M_Z^2)$ of 0.118 is rather larger. Clearly, the freedom to refit the HT corrections with $\alpha_S(M_Z^2)$ is playing a key role in the level of reduction. Comparing the fully free and fixed HT cases more closely,  the former are observed to have rather larger uncertainty bands (as we would expect), which reduces the tension between them, even if this is still clearly present if purely $\Delta \chi^2=1$ uncertainty bands were used. 
At NNLO, shown in Fig.~\ref{fig:as_comps_nnlo}, the impact of HT corrections for the lower $W^2$ cut is rather similar to the aN${}^3$LO  case, again leading to a reduction in the central value.

\section{Summary and Outlook}

In this paper we have presented a range of updates and studies within the MSHT global PDF fit that focus on PDF determination in the high $x$ region, and on improving our understanding of the interplay of various theoretical contributions here. 

We have revisited the question of target mass and higher twist corrections, and presented the result of fits with these included for both the baseline MSHT cuts and a lower $W^2$ cut. This in particular corresponds to the first such analysis at approximate N$^3$LO (aN$^3$LO) in QCD. In order to perform this detailed study we have updated the treatment of some of the fixed-target DIS data included in our fit, replacing the previous structure function determinations, corresponding to an average over beam energies, with 
the cross sections at individual beam energies. This results in better 
control of systematic effects, but is shown to have very small impact on our default PDF sets.
The inclusion of target mass corrections is found to be essential for the lower $W^2$ cut but has a negligible impact for the baseline MSHT case, as we would expect. The impact of higher twist corrections, which we have freely parameterised, is found to be more subtle.  For both cuts, these provide a clear improvement in fit quality but in the baseline case we have demonstrated that their origin can not truly be assigned to a genuine higher twist corrections, but is instead tied up with the missing higher order QCD corrections that are present at higher $x$, effectively absorbing these. 

The PDF impact of these corrections is, on the other hand, relatively mild, with any changes being generally well within PDF uncertainties; this is the case for both the baseline and lower $W^2$ cut. Of particular interest is that this change is found to be rather smaller in the aN$^3$LO case in comparison to NNLO, representing an encouraging trend for increased stability with perturbative order. This result is consistent with the fact that the size of the higher twist corrections are observed to be  smaller at aN$^3$LO. In terms of phenomenology, the inclusion of these corrections is found to reduce the Higgs boson production cross section via gluon--gluon fusion by just below $1\%$ at aN$^3$LO, and about twice this at NNLO.

We have also investigated the impact of these corrections on the preferred value of the strong coupling. We have found that these lead to somewhat of a lowering in the central value, that would even lie close to the edge of the $T^2=1$ uncertainty band of the baseline result. However, with the MSHT dynamic tolerance approach these difference are observed to lie well within uncertainties. 

We have in addition investigated the impact of some new datasets with particular sensitivity to the high $x$ region, namely Seaquest data on fixed--target Drell Yan production, and an alternative presentation of ZEUS inclusive DIS data that extends to the $x\to 1$ region and applies an improved binning. The Seaquest data is found to provide important constraints on the $\overline{d}/\overline{u}$ ratio (or equivalently difference) at high $x$, although a definite tension with the previously used NuSea ratio data is also evident in this region. The impact of the new ZEUS data, which we assess via reweighting, is found to be rather limited, especially when the lower $x$ data are removed in order to avoid double counting with the existing HERA combination data in the fit. 

In summary, we have presented a range of studies of relevance to PDF determination in the high $x$ region. This region is of particular importance for the LHC physics programme, given the sensitivity of high mass BSM searches to it. As such, it is crucial that we have a robust account of all sources of theoretical and experimental uncertainty, with this study being an important further step in that direction.

\section*{Acknowledgements}

L. H.-L. and R. S. T. thank the Science and Technology Facilities
Council (STFC) for support via grant awards ST/T000856/1 and ST/X000516/1. T.C. acknowledges
funding by Research Foundation-Flanders (FWO) (application number: 12E1323N).

\bibliography{references}{}
\bibliographystyle{h-physrev}

\end{document}